\documentclass[prd,aps,nofootinbib,notitlepage,showpacs,preprintnumbers]{revtex4-1}
\usepackage{graphicx,epsf,amsmath,amsfonts,amssymb,amsbsy}
\usepackage{epsfig}
\newcommand{\ds}{\displaystyle}
\newcommand{\vev}[1]{\langle#1\rangle}
\newcommand{\mat}{\left ( \begin{array}}
\newcommand{\emat}{\end{array} \right )}
\newcommand{\vect}{\left ( \begin{array}{c}}
\newcommand{\evect}{\end{array} \right )}

\newcommand{\Det}{\mathop{\rm Det}\nolimits}
\begin{document}

%\hfill HU-EP-11/11
\title{ \bf Duality between chiral symmetry breaking and charged pion condensation
at large $N_c$: Consideration of an NJL$_2$ model with baryon-, isospin- and chiral isospin chemical potentials }

\author{D. Ebert $^{1)}$, T.G. Khunjua $^{2)}$, and K.G. Klimenko $^{3),4)}$}
%, and V.C. Zhukovsky $^{2)}$}
\vspace{1cm}

\affiliation{$^{1)}$ Institute of Physics,
Humboldt-University Berlin, 12489 Berlin, Germany}
\affiliation{$^{2)}$ Faculty of Physics, Moscow State University,
119991, Moscow, Russia} \affiliation{$^{3)}$ State Research Center
of Russian Federation -- Institute for High Energy Physics,
NRC "Kurchatov Institute", 142281, Protvino, Moscow Region, Russia}
\affiliation{$^{4)}$ University ``Dubna`` (Protvino branch), 142281, Protvino, Moscow Region, Russia}

\begin{abstract}
In this paper we investigate the phase structure of a (1+1)-dimensional schematic quark model with four-quark interaction and in the presence of baryon
($\mu_B$), isospin ($\mu_I$) and chiral isospin ($\mu_{I5}$) chemical
potentials. It is established that in the large-$N_c$ limit ($N_c$ is
the number of colored quarks) there exists a duality correspondence between the
chiral symmetry breaking phase and the charged pion condensation (PC) one. The role and influence of this property on the phase structure of the model are studied. Moreover, it is shown that the chemical potential $\mu_{I5}$ promotes the appearance of the charged PC phase with nonzero baryon density.
\end{abstract}
%\pacs{12.39.Ki, 12.38.Mh, 21.65.Qr}
%%% 12.38.Mh Quark-gluon plasma
%%% 21.65.Qr Quark matter
%%% 12.39.Ki Relativistic quark model

\maketitle

\section{Introduction}
%$\dd$

Recently, much attention has been paid to the investigation of
the QCD phase diagram in the presence of baryonic as well as isotopic
(isospin) chemical potentials. The reason is that dense baryonic
matter which can appear in heavy-ion collision experiments has an
evident isospin asymmetry. Moreover, the dense hadronic/quark matter
inside compact stars is also expected to be isotopically asymmetric.
To describe physical situations, when the baryonic density is comparatively low, usually different nonperturbative methods or effective theories such as chiral
effective Lagrangians and especially  Nambu -- Jona-Lasinio (NJL)
type models \cite{njl} are employed. In this way, QCD phase diagrams
including chiral symmetry restoration \cite{asakawa,ebert,sadooghi,hiller,boer}, color superconductivity \cite{alford,klim,incera}, and charged pion condensation (PC) phenomena \cite{son,eklim,ak,mu,andersen,ekkz,gkkz,Mammarella:2015pxa,Carignano:2016lxe} were investigated under heavy-ion experimental and/or compact star conditions, i.e. in the presence of temperature, chemical potentials and possible external (chromo)magnetic fields.

Among all the above mentioned phenomena, which can be observed in dense baryonic matter, the existence of the charged PC phase is predicted without sufficient certainty. Indeed, for some values of model parameters (coupling constant $G$, cutoff parameter $\Lambda$, etc.) the charged PC phase with {\it nonzero baryon density} is allowed by NJL models. However, it is forbidden in the framework of NJL models for other physically interesting values of $G$ and $\Lambda$ \cite{eklim}. Moreover, if the electric charge neutrality constraint is imposed,  the charged pion condensation phenomenon  depends strongly on the bare (current) quark mass
values. In particular, it turns out that the charged PC phase with
{\it nonzero baryonic density} is forbidden in the framework of NJL models, 
if the bare quark masses reach the physically acceptable
values of $5\div 10$ MeV (see ref. \cite{andersen}). Due to these circumstances, the question arises whether there exist factors promoting the appearance of charged PC
phenomenon in  dense baryonic matter. A positive answer to this
question was obtained in the papers \cite{ekkz,gkkz}, where it was
shown that a charged PC phase might be realized in dense baryonic system with finite
size or in the case of a spatially inhomogeneous pion condensate.
These conclusions are demonstrated in \cite{ekkz,gkkz}, using a (1+1)-dimensional
toy model with four-quark interactions and containing baryon and isospin chemical potentials.

In the present paper we will show that a chiral imbalance of dense and isotopically asymmetric baryon matter is another interesting factor, which can induce a charged PC phase. Recall that chiral imbalance, i.e. a nonzero difference between densities of left- and right-handed fermions, may arise from the chiral anomaly in the quark-gluon-plasma phase of QCD and possibly leads to the chiral magnetic effect \cite{fukus} in heavy-ion collisions. It might be realized also in compact stars
or condensed matter systems \cite{andrianov} (see also the review \cite{ms}). Note also that phenomena, connected with a chiral imbalance, are usually described in the framework of NJL models with a chiral chemical potential \cite{andrianov}.

Obviously, the (3+1)-dimensional NJL models depend on the cutoff
parameter which is typically chosen to be of the order of 1 GeV, so that
the results of their usage are valid only at {\it comparatively low
energies, temperatures and densities (chemical potentials)}.
Moreover, there exists also a class of renormalizable theories, the
(1+1)-dimensional chiral Gross--Neveu (GN) type models \cite{gn,ft},
\footnote{Below we shall use the notation ``NJL$_2$ model''  instead
of ``chiral GN model'' for (1+1)-dimensional models with {\it a
continuous chiral and/or isotopic, etc, symmetries}, since the chiral structure of the Lagrangian is the same as that of the corresponding (3+1)-dimensional NJL model.} that can be used as a laboratory for the qualitative simulation of
specific properties of QCD at {\it arbitrary energies}.
Renormalizability, asymptotic freedom, as well as the spontaneous
breaking of chiral symmetry (in vacuum) are the most fundamental
inherent features both for QCD and all NJL$_2$ type models. In addition,
the $\mu_B-T$ phase diagram (with $\mu_B$ the baryon number chemical potential and $T$ the temperature) is qualitatively the same for the QCD and NJL$_2$ models
\cite{wolff,kgk1,barducci,chodos}. Let us further mention that
(1+1)-dimensional Gross-Neveu type models are also suitable for the description of physics in quasi one-dimensional condensed matter systems like polyacetylene
\cite{caldas}. It is currently well understood (see, e.g., the
discussion in \cite{barducci,chodos,thies}) that the usual {\it no-go}
theorem \cite{coleman}, which generally forbids the spontaneous
breaking of any continuous symmetry in two-dimensional spacetime,
does not work in the limit  $N_c\to\infty$, where $N_c$ is the
number of colored quarks. This follows directly from the fact that in the
limit of large $N_c$ the quantum fluctuations, which would otherwise
destroy a long-range order corresponding to a spontaneous symmetry
breaking, are suppressed by $1/N_c$ factors. Thus,  the effects
inherent for real dense quark matter, such as chiral symmetry breaking
phenomenon  (spontaneous breaking of the continuous axial $U(1)$
symmetry) or charged pion condensation (spontaneous breaking of the
continuous isospin symmetry) might be simulated in terms of a simpler
(1+1)-dimensional NJL-type model, though only in the leading order of
the large $N_c$ approximation (see, e.g., Refs. \cite{thies} and \cite{ektz,massive,ek2,gubina}, respectively).

This paper is devoted to the investigation of the charged PC
phenomenon in the framework of an extended (1+1)-dimensional NJL model with two quark flavors and in the presence of the baryon ($\mu_B$), isospin ($\mu_I$)  as well as chiral isospin ($\mu_{I5}$) chemical potentials. Moreover, as usual, it is convenient to perform all calculations in the leading order of the large $N_c$ technique. In order to clarify the true role of the chiral isospin chemical potential $\mu_{I5}$ in the creation of the charged PC in dense quark matter, we suppose throughout the paper that all condensates are spatially homogeneous.\footnote{As it was noted above, spatial inhomogeneity of condensates by itself can cause charged PC in dense baryon matter \cite{gkkz}.} Under this constraint the model was already investigated earlier at $\mu_{I5}=0$ \cite{ektz,massive,ek2}, where it was shown that the charged PC phase with nonzero baryon density is forbidden at arbitrary values of $\mu_B$ and $\mu_I$. In contrast, we show that at $\mu_{I5}\ne 0$, i.e. when there is an isotopic chiral imbalance of the system, the charged PC phase with nonzero baryon density is allowed to exist. This fact, i.e. the promotion of the charged PC phenomenon in dense quark/baryon matter by nonzero values of $\mu_{I5}$, is the main result of the present paper. In addiion, we show that in the leading order of the large-$N_c$ approximation there arises a duality between chiral symmetry breaking (CSB) and charged PC phenomena in the framework of the NJL$_2$ model under consideration. It means that if at $\mu_I=A$ and $\mu_{I5}=B$ (at arbitrary fixed chemical potential $\mu$), e.g., the CSB (or the
charged PC) phase is realized in the model, then at the permuted
values of  these chemical potentials, i.e. at $\mu_I=B$ and
$\mu_{I5}=A$, the charged PC (or the CSB) phase is arranged. So, it is
enough to know the phase structure of the model at $\mu_I<\mu_{I5}$,
in order to establish the phase structure at $\mu_I>\mu_{I5}$. Knowing condensates
and other dynamical and thermodynamical quantities of the system,
e.g. in the CSB phase, one can then obtain the corresponding
quantities in the dually conjugated charged PC  phase of the model, by
simply performing there the duality transformation,
$\mu_I\leftrightarrow\mu_{I5}$. \footnote{Note that another kind of 
duality correspondence, the duality between CSB and superconductivity, was demonstrated both in (1+1)- and (2+1)-dimensional NJL models \cite{thies2,ekkz2}. }

The paper is organized as follows. In Sec. II a toy (1+1)-dimensional NJL-type model
with two quark flavors ($u$ and $d$ quarks) and including
three kinds of chemical potentials, $\mu_B,\mu_I,\mu_{I5}$, is presented. 
Next, the unrenormalized thermodynamic potential (TDP) of the
NJL$_2$-type model is given in the leading order of the large-$N_c$
expansion. Here the dual symmetry of the model TDP is
established. It means that it is invariant under the simultaneous interchange of
$\mu_I,\mu_{I5}$ chemical potentials and chiral and
charged pion condensates. In Sec. III the renormalization of the
TDP is performed. Sec. IV contains a detailed numerical investigation
of various  phase portraits with particular emphasis on the role of
the duality symmetry of the TDP. It is clear from this consideration
that in the framework of our model the charged PC phenomenon of dense
and isotopically asymmetric quark matter is allowed only if in addition 
there is a chiral isotopic asymmetry of matter, i.e. in the case $\mu_{I5}\ne 0$. Some technical details are relegated to Appendices A and B. 

\section{ The model and its thermodynamic potential}

We consider a (1+1)-dimensional NJL model in order to mimic the phase structure of real dense quark matter with two massless quark flavors ($u$ and $d$ quarks). Its Lagrangian, which is symmetrical under global color SU($N_c$) group, has the form
\begin{eqnarray}
&&  L=\bar q\Big [\gamma^\nu\mathrm{i}\partial_\nu
+\frac{\mu_B}{3}\gamma^0+\frac{\mu_I}2 \tau_3\gamma^0+\frac{\mu_{I5}}2 \tau_3\gamma^0\gamma^5\Big ]q+ \frac
{G}{N_c}\Big [(\bar qq)^2+(\bar q\mathrm{i}\gamma^5\vec\tau q)^2 \Big
],  \label{1}
\end{eqnarray}
where the quark field $q(x)\equiv q_{i\alpha}(x)$ is a flavor doublet
($i=1,2$ or $i=u,d$) and color $N_c$-plet ($\alpha=1,...,N_c$) as
well as a two-component Dirac spinor (the summation in (\ref{1})
over flavor, color, and spinor indices is implied); $\tau_k$
($k=1,2,3$) are Pauli matrices in two-dimensional flavor space. The Dirac $\gamma^\nu$-matrices ($\nu=0,1$) and $\gamma^5$ in (1) are matrices in 
two-dimensional spinor space,
\begin{equation}
\begin{split}
\gamma^0=
\begin{pmatrix}
0&1\\
1&0\\
\end{pmatrix};\qquad
\gamma^1=
\begin{pmatrix}
0&-1\\
1&0\\
\end{pmatrix};\qquad
\gamma^5=\gamma^0\gamma^1=
\begin{pmatrix}
1&0\\
0&{-1}\\
\end{pmatrix}.
\end{split}
\end{equation}
Note that at $\mu_{I5}=0$ the model was already investigated in details, e.g., in Refs \cite{ektz,massive,ek2,gubina}. It is evident that the model (\ref{1}) is a generalization of the two-dimensional GN model \cite{gn} with a single massless quark color $N_c$-plet to the case of two quark flavors
and additional baryon $\mu_B$-, isospin $\mu_I$- and axial isospin $\mu_{I5}$ chemical potentials. These parameters are introduced in order to describe in the framework of the model (1) quark matter with nonzero baryon $n_B$-, isospin $n_I$- and axial isospin $n_{I5}$ densities, respectively.
It is evident that Lagrangian (1), both at $\mu_{I5}=0$ and $\mu_{I5}\ne 0$, is invariant with respect to the abelian $U_B(1)$, $U_{I_3}(1)$ and $U_{AI_3}(1)$ groups, where \footnote{\label{f1,1}
Recall for the following that~~
$\exp (\mathrm{i}\beta\tau_3/2)=\cos (\beta/2)
+\mathrm{i}\tau_3\sin (\beta/2)$,~~~~
$\exp (\mathrm{i}\omega\gamma^5\tau_3/2)=\cos (\omega/2)
+\mathrm{i}\gamma^5\tau_3\sin (\omega/2)$.}
\begin{eqnarray}
U_B(1):~q\to\exp (\mathrm{i}\alpha/3) q;~
U_{I_3}(1):~q\to\exp (\mathrm{i}\beta\tau_3/2) q;~
U_{AI_3}(1):~q\to\exp (\mathrm{i}
\omega\gamma^5\tau_3/2) q.
\label{2001}
\end{eqnarray}
(In (\ref{2001}) the real parameters $\alpha,\beta,\omega$ specify an
arbitrary element of the $U_B(1)$, $U_{I_3}(1)$ and $U_{AI_3}(1)$ groups, respectively.) So the quark bilinears $\frac 13\bar q\gamma^0q$, $\frac 12\bar q\gamma^0\tau^3 q$ and $\frac 12\bar q\gamma^0\gamma^5\tau^3 q$ are the zero components of corresponding conserved currents. Their ground state expectation values are just the baryon $n_B$-, isospin $n_I$- and chiral (axial) isospin $n_{I5}$ densities of quark matter, i.e. $n_B=\frac 13\vev{\bar q\gamma^0q}$, $n_I=\frac 12\vev{\bar q\gamma^0\tau^3 q}$
and $n_{I5}=\frac 12\vev{\bar q\gamma^0\gamma^5\tau^3 q}$. As usual, the quantities $n_B$, $n_I$ and $n_{I5}$ can be also found by differentiating the thermodynamic potential of the system with respect to the corresponding chemical potentials. The goal of the
present paper is the investigation of the ground state properties and phase structure of the system (1) and its dependence on the chemical potentials  $\mu_B$, $\mu_I$ and $\mu_{I5}$.

To find the thermodynamic potential of the system, we use a semi-bosonized version of the Lagrangian (\ref{1}), which contains composite bosonic fields $\sigma (x)$ and $\pi_a (x)$ $(a=1,2,3)$ (in what follows, we use the notations $\mu\equiv\mu_B/3$, $\nu=\mu_I/2$ and $\nu_{5}=\mu_{I5}/2$):
\begin{eqnarray}
\widetilde L\ds &=&\bar q\Big [\gamma^\rho\mathrm{i}\partial_\rho
+\mu\gamma^0 + \nu\tau_3\gamma^0+\nu_{5}\tau_3\gamma^0\gamma^5-\sigma
-\mathrm{i}\gamma^5\pi_a\tau_a\Big ]q
 -\frac{N_c}{4G}\Big [\sigma\sigma+\pi_a\pi_a\Big ].
\label{2}
\end{eqnarray}
In (\ref{2}) the summation over repeated indices is implied.
From the Lagrangian (\ref{2}) one gets the Euler--Lagrange equations
for the bosonic fields
\begin{eqnarray}
\sigma(x)=-2\frac G{N_c}(\bar qq);~~~\pi_a (x)=-2\frac G{N_c}(\bar q
\mathrm{i}\gamma^5\tau_a q).
\label{200}
\end{eqnarray}
Note that the composite bosonic field $\pi_3 (x)$ can be identified
with the physical $\pi_0$ meson, whereas the physical $\pi^\pm
(x)$-meson fields are the following combinations of the composite
fields, 
%%%%%%% I remember opposite signs for $\pi^\pm$ :  \pi_1 (x) -/+ i\pi_2 (x). Check!
$\pi^\pm (x)=(\pi_1 (x)\pm i\pi_2 (x))/\sqrt{2}$. 
%%%%%%%%%%%%%%%%%%%%%%%%%%%%%%%%%%%%%%%%%%%%%%%%%%%%%%%%%%%%%%%
Obviously, the semi-bosonized Lagrangian $\widetilde L$ is equivalent to the initial Lagrangian (\ref{1}) when using the equations (\ref{200}). Furthermore, it is clear from (\ref{2001}), (\ref{200}) and footnote \ref{f1,1} that the bosonic fields transform under the isospin $U_{I_3}(1)$ and axial isospin $U_{AI_3}(1)$ groups in the following manner:
\begin{eqnarray}
U_{I_3}(1):~&&\sigma\to\sigma;~~\pi_3\to\pi_3;~~\pi_1\to\cos
(\beta)\pi_1+\sin (\beta)\pi_2;~~\pi_2\to\cos (\beta)\pi_2-\sin
(\beta)\pi_1,\nonumber\\
U_{AI_3}(1):~&&\pi_1\to\pi_1;~~\pi_2\to\pi_2;~~\sigma\to\cos
(\omega)\sigma+\sin (\omega)\pi_3;~~\pi_3\to\cos
(\omega)\pi_3-\sin (\omega)\sigma.
\label{201}
\end{eqnarray}
In general the phase structure of a given model is characterized by the behaviour of some quantities, called order parameters (or condensates), vs external conditions (temperature, chemical potentials, etc). In the case of model (1), such order parameters are the ground state expectation values of the composite fields, i.e. the quantities $\vev{\sigma (x)}$ and $\vev{\pi_a (x)}$ $(a=1,2,3)$. It is clear from (\ref{201}) that if $\vev{\sigma(x)}\ne 0$ and/or $\vev{\pi_3(x)}\ne 0$, then the axial isospin $U_{AI_3}(1)$ symmetry of the model is spontaneously broken down, whereas at $\vev{\pi_1(x)}\ne 0$ and/or $\vev{\pi_2(x)}\ne 0$ we have a spontaneous breaking of the isospin $U_{I_3}(1)$ symmetry. Since in the last case the ground state expectation values (condensates) of both the fields $\pi^+(x)$ and $\pi^-(x)$ are not zero, this phase is usually called charged pion condensation (PC) phase. The ground state expectation values $\vev{\sigma(x)}$ and $\vev{\pi_a(x)}$ are the coordinates of the global minimum point of the thermodynamic potential $\Omega (\sigma,\pi_a)$ of the system.

Starting from the theory (\ref{2}), one obtains in the leading order
of the large $N_c$-expansion (i.e. in the one-fermion loop
approximation) the following path integral expression for the
effective action ${\cal S}_{\rm {eff}}(\sigma,\pi_a)$ of the bosonic
$\sigma (x)$ and $\pi_a (x)$ fields:
$$
\exp(\mathrm{i}{\cal S}_{\rm {eff}}(\sigma,\pi_a))=
  N'\int[d\bar q][dq]\exp\Bigl(\mathrm{i}\int\widetilde L\,d^2 x\Bigr),
$$
where
\begin{equation}
{\cal S}_{\rm {eff}}(\sigma,\pi_a)
=-N_c\int d^2x\left [\frac{\sigma^2+\pi^2_a}{4G}
\right ]+\widetilde {\cal S}_{\rm {eff}},
\label{3}
\end{equation}
$N'$ is a normalization constant. The quark contribution to the effective action, i.e. the term $\widetilde {\cal S}_{\rm {eff}}$ in (\ref{3}), is given by:
\begin{equation}
\exp(\mathrm{i}\widetilde {\cal S}_{\rm {eff}})=N'\int [d\bar
q][dq]\exp\Bigl(\mathrm{i}\int\Big [\bar q\mathrm{D}q\Big ]d^4
x\Bigr)=[\Det D]^{N_c}.
 \label{4}
\end{equation}
In (\ref{4}) we have used the notation $\mathrm{D}\equiv D\times
\mathrm{I}_c$, where $\mathrm{I}_c$ is the unit operator in the
$N_c$-dimensional color space and
\begin{equation}
D\equiv\gamma^\nu\mathrm{i}\partial_\nu +\mu\gamma^0
+ \nu\tau_3\gamma^0+\nu_{5}\tau_3\gamma^0\gamma^5-\sigma -\mathrm{i}\gamma^5\pi_a\tau_a
\label{5}
\end{equation}
is the Dirac operator, which acts in the flavor-, spinor- as well as
coordinate spaces only. Using the general formula $\Det D=\exp {\rm
Tr}\ln D$, one obtains for the effective action the following expression
\begin{equation}
{\cal S}_{\rm {eff}}(\sigma,\pi_a)=-N_c\int
d^2x\left[\frac{\sigma^2+\pi^2_a}{4G}\right]-\mathrm{i}N_c{\rm
Tr}_{sfx}\ln D,
\label{6}
\end{equation}
where the Tr-operation stands for the trace in spinor- ($s$), flavor-
($f$) as well as two-dimensional coordinate- ($x$) spaces,
respectively. Using (\ref{6}), we obtain the thermodynamic
potential (TDP) $\Omega (\sigma,\pi_a)$ of the system:
\begin{eqnarray}
\Omega (\sigma,\pi_a)~
&&\equiv -\frac{{\cal S}_{\rm {eff}}(\sigma,\pi_a)}{N_c\int
d^2x}~\bigg |_{~\sigma,\pi_a=\rm {const}}
=\frac{\sigma^2+\pi^2_a}{4G}+\mathrm{i}\frac{{\rm
Tr}_{sfx}\ln D}{\int d^2x}\nonumber\\
&&=\frac{\sigma^2+\pi^2_a}{4G}+\mathrm{i}{\rm
Tr}_{sf}\int\frac{d^2p}{(2\pi)^2}\ln\overline{D}(p),
\label{7}
\end{eqnarray}
where the $\sigma$- and $\pi_a$ fields are now $x$-independent quantities, and
\begin{equation}
\overline{D}(p)=\not\!p +\mu\gamma^0
+ \nu\tau_3\gamma^0+ \nu_{5}\tau_3\gamma^0\gamma^5-\sigma
-\mathrm{i}\gamma^5\pi_a\tau_a
\label{50}
\end{equation}
is the momentum space representation of the Dirac operator $D$ (\ref{5}). In what follows we are going to investigate the $\mu,\nu,\nu_{5}$-dependence of the global minimum point of the function $\Omega (\sigma,\pi_a)$ vs $\sigma,\pi_a$. To simplify the task, let us note that due to the $U_{I_3}(1)\times U_{AI_3}(1)$ invariance of the model, the TDP (\ref{7}) depends effectively only on the two combinations $\sigma^2+\pi_3^2$ and $\pi_1^2+\pi_2^2$ of the bosonic fields, as is easily seen from (\ref{201}). In this case, without loss of generality, one can put $\pi_2=\pi_3=0$ in (\ref{7}), and study the TDP (\ref{7}) as a function of only two variables, $M\equiv\sigma$ and $\Delta\equiv\pi_1$. Taking into account this constraint in (\ref{50}) and (\ref{7}) as well as the general relation
\begin{eqnarray}
{\rm Tr}_{sf}\ln\overline{D}(p)=\ln\Det\overline{D}(p)=\sum_i\ln\epsilon_i,
\end{eqnarray}
where the summation over all four eigenvalues $\epsilon_i$ of the 4$\times$4 matrix $\overline{D}(p)$ is implied and
\begin{eqnarray}
\epsilon_{1,2,3,4}=-M\pm\sqrt{(p_0+\mu)^2-p_1^2-\Delta^2+
\nu^2-\nu_{5}^2\pm 2\sqrt{\big [(p_0+\mu)\nu+p_1\nu_{5}\big ]^2-\Delta^2(\nu^2-\nu_{5}^2)}},
\label{8}
\end{eqnarray}
we have from (\ref{7})
\begin{eqnarray}
\Omega (M,\Delta)~&&
=\frac{M^2+\Delta^2}{4G}+\mathrm{i}\int\frac{d^2p}{(2\pi)^2}\ln
P_4(p_0).
\label{9}
\end{eqnarray}
In (\ref{9}) we use the notations
\begin{eqnarray}
P_4(p_0)=\epsilon_1\epsilon_2\epsilon_3\epsilon_4=\eta^4-2a\eta^2-b\eta+c,
\label{91}
\end{eqnarray}
where $\eta=p_0+\mu$ and
\begin{eqnarray}
a&&=M^2+\Delta^2+p_1^2+\nu^2+\nu_{5}^2;~~b=8p_1\nu\nu_{5};\nonumber\\
c&&=a^2-4p_1^2(\nu^2+\nu_5^2)-4M^2\nu^2-4\Delta^2\nu_5^2-4\nu^2\nu_5^2.
\label{10}
\end{eqnarray}
It is evident from (\ref{10}) that the TDP (\ref{9}) is an even
function over each of the variables $M$ and $\Delta$.
In addition, it is invariant under each of the transformations
$\mu\to-\mu$,  $\nu\to-\nu$, $\nu_5\to-\nu_5$.
\footnote{Indeed, if we perform simultaneously with $\mu\to-\mu$
the change of variables $p_0\to-p_0$ and $p_1\to-p_1$ in the integral (\ref{9}),
then one can easily see that the expression (\ref{9}) remains
intact. Finally, if only $\nu$ (only $\nu_5$) is replaced by $-\nu$
(by $-\nu_5$), we should transform $p_1\to-p_1$ in the integral (\ref{9}) in order to see that the TDP remains unchanged. } Hence, without loss of generality we can consider in the following only $\mu\ge 0$, $\nu\ge 0$, $\nu_5\ge 0$, $M\ge 0$, and $\Delta\ge 0$ values of these quantities. In powers of $\Delta$ the fourth-degree polynomial $P_4(p_0)$ has the following form
\begin{eqnarray}
P_4(p_0)&\equiv&\Delta^4-2\Delta^2(\eta^2-p_1^2-M^2+\nu_5^2-\nu^2)\nonumber\\
&+&\big [M^2+(p_1-\nu_5)^2-(\eta+\nu)^2\big ]\big [M^2+(p_1+\nu_5)^2-
(\eta-\nu)^2\big ]. \label{17}
\end{eqnarray}
Expanding the right-hand side of (\ref{17}) in powers of $M$, one can
obtain an equivalent alternative expression for this polynomial. Namely,
\begin{eqnarray}
P_4(p_0)&\equiv& M^4-2M^2(\eta^2-p_1^2-\Delta^2+\nu^2-\nu_5^2)\nonumber\\
&+&\big [\Delta^2+(p_1-\nu)^2-(\eta+\nu_5)^2\big ]\big
[\Delta^2+(p_1+\nu)^2-(\eta-\nu_5)^2\big ].\label{18}
\end{eqnarray}
Thus, we find that the TDP (\ref{9}) is invariant with respect to the so-called duality transformation (for an analogous case of duality between chiral and superconducting condensates, see \cite{thies2,ekkz2}),
\begin{eqnarray}
{\cal D}:~~~~M\longleftrightarrow \Delta,~~\nu\longleftrightarrow\nu_5.
 \label{16}
\end{eqnarray}
Note that according to the general theorem of algebra, the polynomial
$P_4(p_0)$ can be presented also in the form
\begin{eqnarray}
P_4(p_0)\equiv (p_0-p_{01})(p_0-p_{02})(p_0-p_{03})(p_0-p_{04}), \label{170}
\end{eqnarray}
where the roots $p_{01}$, $p_{02}$, $p_{03}$ and $p_{04}$ of this polynomial 
are the energies of quasiparticle or quasiantiparticle excitations of the system. In particular, it follows from (\ref{17}) that at $\Delta=0$ the set of roots $p_{0i}$ looks like
\begin{eqnarray}
\Big\{p_{01},p_{02},p_{03},p_{04}\Big\}\Big |_{\Delta=0}=\Big\{-\mu-\nu\pm\sqrt{M^2+(p_1-\nu_5)^2},-\mu+\nu\pm\sqrt{M^2+(p_1+\nu_5)^2}\Big\}, \label{26}
\end{eqnarray}
whereas it is clear from (\ref{18}) that at $M=0$ it has the form
\begin{eqnarray}
\Big\{p_{01},p_{02},p_{03},p_{04}\Big\}\Big |_{M=0}=\Big\{-\mu-\nu_5\pm\sqrt{\Delta^2+(p_1-\nu)^2},-\mu+\nu_5\pm\sqrt{\Delta^2+(p_1+\nu)^2}\Big\}. \label{27}
\end{eqnarray}
Taking into account the relation (\ref{170}) as well as the formula (\cite{gkkz})
\begin{eqnarray}
\int_{-\infty}^\infty dp_0\ln\big
(p_0-K)=\mathrm{i}\pi|K|,\label{int}
\end{eqnarray}
(being true up to an infinite term independent of the real quantity $K$),
it is possible to integrate in (\ref{9}) over $p_0$. Then, the {\it unrenormalized}
TDP (\ref{9}) can be presented in the following form,
\begin{eqnarray}
\Omega (M,\Delta)&\equiv&\Omega^{un} (M,\Delta)=
\frac{M^2+\Delta^2}{4G}-
\int_{-\infty}^\infty\frac{dp_1}{4\pi}\Big (|p_{01}|+|p_{02}|+|p_{03}|+|p_{04}|\Big ). \label{28}
\end{eqnarray}

\section{Calculation of the TDP}
\subsection{Thermodynamic potential in the vacuum case: $\mu=0,\nu=0,\nu_5=0$}

First of all, let us obtain a finite, i.e. renormalized, expression
for the TDP (\ref{28}) at $\mu=0$, $\nu=0$ and $\nu_5=0$, i.e. in
vacuum. Since in this case a thermodynamic potential is usually called
effective potential, we use for it the notation $V^{un}
(M,\Delta)$. As a consequence of (\ref{9})-(\ref{10}) and using (\ref{int}),
it is clear that at $\mu=\nu=\nu_5=0$ the effective potential $V^{un} (M,\Delta)$ looks like
\begin{eqnarray}
V^{un} (M,\Delta)&=&
\frac{M^2+\Delta^2}{4G}
+2i\int\frac{d^2p}{(2\pi)^2}\ln\Big
[p_0^2-p_1^2-M^2-\Delta^2\Big ]\nonumber\\
&=&\frac{M^2+\Delta^2}{4G}-\int_{-\infty}^\infty\frac{dp_1}{\pi}\sqrt{p_1^2+M^2+\Delta^2}. \label{25}
\end{eqnarray}
It is evident that the effective potential (\ref{25}) is an ultraviolet divergent
quantity. So, we need to renormalize it. This procedure consists of
two steps: (i) First of all we need to regularize the divergent integral in (\ref{25}), i.e. we suppose there that $|p_1|<\Lambda$. (ii)  Second, we must suppose also that the bare coupling constant $G$ depends on the cutoff parameter
$\Lambda$ in such a way that in the limit $\Lambda\to\infty$ one obtains a finite expression for the effective potential.

Following the step (i) of this procedure, we have
\begin{eqnarray}
V^{reg} (M,\Delta)&=&\frac{M^2+\Delta^2}{4G}-\frac 2\pi\int_{0}^\Lambda dp_1\sqrt{p_1^2+M^2+\Delta^2}\nonumber\\
=\frac{M^2+\Delta^2}{4G}&-&\frac 1\pi\left\{\Lambda\sqrt{\Lambda^2+M^2+\Delta^2}+(M^2+\Delta^2)\ln\frac{\Lambda+\sqrt{\Lambda^2+M^2+\Delta^2}}{\sqrt{M^2+\Delta^2}}\right\}. \label{29}
\end{eqnarray}
Further, according to step (ii) we suppose that in (\ref{29}) the bare coupling $G\equiv G(\Lambda)$ has the following $\Lambda$ dependence:
\begin{eqnarray}
\frac 1{4G(\Lambda)}=\frac 1\pi\ln\frac{2\Lambda}{m}, \label{30}
\end{eqnarray}
where $m$ is a new free mass scale of the model, which appears instead of the dimensionless bare coupling constant $G$ (dimensional transmutation) and, evidently, does not depend on a normalization point, i.e. it is a renormalization
invariant quantity. Substituting (\ref{30}) into (\ref{29}) and
ignoring there an unessential term $(-\Lambda^2/\pi)$, we obtain
in the limit $\Lambda\to\infty$ the finite and renormalization invariant expression for the effective potential,
\begin{eqnarray}
V_0 (M,\Delta)&=&\frac{M^2+\Delta^2}{2\pi}\left [\ln\left (\frac{M^2+\Delta^2}{m^2}\right )-1\right ]. \label{31}
\end{eqnarray}

\subsection{Calculation of the TDP (\ref{28}) in the general case: $\mu>0$, $\nu>0$, $\nu_5>0$}

In Appendix \ref{ApB} the properties of the quasiparticle energies
$p_{0i}$, where $i=1,...,4$, are investigated. In particular, it is
clear from the asymptotic expansion (\ref{B9}) that the integral
over $p_1$ in (\ref{28}) is ultraviolet divergent. Since the asymptotic expansion (\ref{B9}) does not depend on chemical potentials $\mu$, $\nu$ and $\nu_5$, one can transform the expression (\ref{28}) in the following way,
\begin{eqnarray}
\Omega^{un} (M,\Delta)&=&
\frac{M^2+\Delta^2}{4G}-
\int_{0}^\infty\frac{dp_1}{2\pi}\Big (|p_{01}|+|p_{02}|+|p_{03}|+|p_{04}|\Big )\Big |_{\mu=\nu=\nu_5=0}\nonumber\\
&-&\int_{0}^\infty\frac{dp_1}{2\pi}\Big [\sum_{i=1}^4|p_{0i}|-\Big (\sum_{i=1}^4|p_{0i}|\Big )\Big |_{\mu=\nu=\nu_5=0}\Big ], \label{32}
\end{eqnarray}
where we took into account that the quantities $p_{0i}$ are even
functions with respect to $p_1$ (see Appendix \ref{ApB}). Now it is
evident that the last integral in (\ref{32}) is convergent and all
ulraviolet divergences of the TDP are located in the first integral of
(\ref{32}). Moreover, it is clear due to the relation (\ref{B10}) that
the first two terms in the right hand side of Eq. (\ref{32}) are just
the unrenormalized effective potential in vacuum (\ref{25}). So to
obtain a finite expression for the TDP (\ref{32}), it is enough to proceed as in the
previous subsection, where just these two terms, i.e. the vacuum
effective potential, were renormalized. As a result, we have
\begin{eqnarray}
\Omega^{ren}
(M,\Delta)&=&V_0(M,\Delta)-\int^\infty_{0}\frac{dp_1}{2\pi}\Big\{|p_{01}|+|p_{02}|+|p_{03}|+|p_{04}|-4\sqrt{p_1^2+M^2+\Delta^2}\Big\}, \label{35}
\end{eqnarray}
where $V_0(M,\Delta)$ is the renormalized TDP (effective potential) (\ref{31}) of the model at $\mu=\nu=\mu_5=0$. Moreover, we have used in (\ref{35}) the relation (\ref{B10}) for the sum of quasiparticle energies in vacuum. Note also that (as it follows from the considerations of Appendix A) the quasiparticle energies $p_{0i}$, where $i=1,...,4$, are invariant (up to a possible permutation of their values) with respect to the duality transformation (\ref{16}). So the renormalized TDP (\ref{35}) is also symmetric under the duality transformation $\cal D$.

Let us denote by $(M_0,\Delta_0)$ the global minimum point (GMP) of the TDP (\ref{35}). Then, investigating the behavior of this point vs $\mu$, $\nu$ and $\nu_5$ it is possible to construct the $(\mu,\nu,\nu_5)$-phase portrait (diagram) of the model. A numerical algorithm for finding the quasi(anti)particle energies  $p_{01}$, $p_{02}$, $p_{03}$, and $p_{04}$ is elaborated in Appendix \ref{ApB}. Based on this, it can be shown numerically that the
GMP of the TDP can never be of the form $(M_0\ne 0,\Delta_0\ne 0)$. Hence, in order to establish the phase portrait of the model, it is enough to study the projections $F_1(M)\equiv\Omega^{ren} (M,\Delta=0)$ and $F_2(\Delta)\equiv\Omega^{ren}(M=0,\Delta)$ of the TDP (\ref{35}) to the $M$ and $\Delta$ axes, correspondingly. Taking into account the relations (\ref{26}) and (\ref{27}) for the quasiparticle energies $p_{0i}$  at $\Delta=0$ or $M=0$, it is possible  to obtain the following expressions for these quantities,
\begin{eqnarray}
F_1(M)&=&\frac{M^2}{2\pi}\ln\left
(\frac{M^2}{m^2}\right )-\frac{M^2}{2\pi}-\frac{\nu_5^2}{\pi}-\frac{\theta (|\mu-\nu|-M)\theta (\sqrt{(\mu-\nu)^2-M^2}-\nu_5)}{2\pi}\left
(|\mu-\nu|\sqrt{(\mu-\nu)^2-M^2}\right.\nonumber\\
&+&\left.\nu_5\sqrt{\nu_5^2+M^2}-2|\mu-\nu|\nu_5-M^2\ln\frac{|\mu-\nu|+\sqrt{|\mu-\nu|^2-M^2}}{\nu_5+\sqrt{\nu_5^2+M^2}}\right )-\frac{\theta (\mu+\nu-M)}{\pi}\left ((\mu+\nu)\sqrt{(\mu+\nu)^2-M^2}\right.\nonumber\\
&-&\left.M^2\ln\frac{\mu+\nu+\sqrt{(\mu+\nu)^2-M^2}}{M}\right )+\frac{\theta (\mu+\nu-M)\theta (\sqrt{(\mu+\nu)^2-M^2}-\nu_5)}{2\pi}\left
((\mu+\nu)\sqrt{(\mu+\nu)^2-M^2}\right.\nonumber\\
&+&\left.\nu_5\sqrt{\nu_5^2+M^2}-2(\mu+\nu)\nu_5-M^2\ln\frac{\mu+\nu+\sqrt{(\mu+\nu)^2-M^2}}{\nu_5+\sqrt{\nu_5^2+M^2}}\right ),\label{33}
\end{eqnarray}
\begin{eqnarray}
F_2(\Delta)&=&F_1(\Delta)\Bigg |_{\nu\longleftrightarrow\nu_5}. \label{34}
\end{eqnarray}
(Details of the derivation of these expressions are given in Appendix
\ref{ApD}.) After simple transformations, one can see that $F_1(M)$ and $F_2(\Delta)$ coincide at $\nu_5=0$ with corresponding TDPs (12) and (13) of the paper \cite{ek2}.

Moreover, it is obvious that the global minimum point of the TDP (\ref{35}) is defined by a comparison between the least values of the functions $F_1(M)$ and $F_2(\Delta)$.

\subsection{Quark number density}

As it is clear from the above consideration, there are three phases in
the model (1).  The first one is the symmetric phase, which
corresponds to the global minimum point $(M_0,\Delta_0)$ of the TDP
(\ref{35}) of the form $(M_0=0,\Delta_0=0)$. In the CSB phase the TDP
reaches the least value at the point $(M_0\ne 0,\Delta_0=0)$. Finally,
in the charged PC phase the global minimum point lies at the point $(M_0=0,\Delta_0\ne 0)$. (Notice, that in the most general case the coordinates (condensates) $M_0$ and $\Delta_0$ of the global minimum point depend on chemical potentials.)

In the present subsection we would like to obtain the expression for the quark number (or particle) density $n_q$ in the ground state of each phase. Recall that in the most general case this quantity is defined by the relation \footnote{The density of baryons $n_B$ and the quark number density $n_q$ are connected by the relation $n_q=3n_B$.}
\begin{eqnarray}
n_q=-\frac{\partial\Omega^{ren}(M_0,\Delta_0)}{\partial\mu}. \label{37}
\end{eqnarray}
Hence, in the chiral symmetry breaking phase we have
\begin{eqnarray}
n_q\bigg |_{CSB}&=&-\frac{\partial\Omega^{ren}(M_0\ne 0,\Delta_0=0)}{\partial\mu}=-\frac{\partial F_1(M_0)}{\partial\mu}=\frac{2\theta\left (\mu+\nu-M_0\right )}{\pi}\sqrt{(\mu+\nu)^2-M_0^2}\nonumber\\
&-&\frac{\theta\left (\mu+\nu-M_0\right )\theta\left (\sqrt{(\mu+\nu)^2-M_0^2}-\nu_5\right )}{\pi}\left [\sqrt{(\mu+\nu)^2-M_0^2}-\nu_5\right ]\nonumber\\
&+&\frac{{\rm sign}(\mu-\nu)\theta\left (|\mu-\nu|-M_0\right )\theta\left (\sqrt{(\mu-\nu)^2-M_0^2}-\nu_5\right )}{\pi}\left [\sqrt{(\mu-\nu)^2-M_0^2}-\nu_5\right ], \label{38}
\end{eqnarray}
where ${\rm sign(x)}$ denotes the sign function and the quantity $F_1(M)$ is given in (\ref{33}).
The quark number density in the charged pion condensation phase can be
easily obtained from (\ref{38}) by the simple replacement,
\begin{eqnarray}
n_q\bigg |_{PC}=-\frac{\partial\Omega^{ren}(M_0=0,\Delta_0\ne 0)}{\partial\mu}=-\frac{\partial F_2(\Delta_0)}{\partial\mu}=\left\{n_q\big |_{CSB}\right\}\bigg |_{M_0\to\Delta_0;~\nu\longleftrightarrow\nu_5}, \label{39}\end{eqnarray}
which is due to the relation (\ref{34}). Supposing in (\ref{38}) that $M_0=0$ and using there the general relation $\theta (x)+\theta(-x)=1$, one can find the following expression for the particle density in the symmetric phase (of course, we take into account the constraints $\mu\ge 0$, $\nu\ge 0$ and $\nu_5\ge 0$)
\begin{eqnarray}
n_q\bigg |_{SYM}&=&\frac{\mu+\nu+\nu_5}{\pi}-\frac{\theta(\nu_5-\mu-\nu)}{\pi}(\nu_5-\mu-\nu)%\nonumber\\&+&
+\frac{{\rm sign}(\mu-\nu)\theta\left (|\mu-\nu|-\nu_5\right )}{\pi}(|\mu-\nu|-\nu_5). \label{40}
\end{eqnarray}
Alternatively, one can find the expression for $n_q\bigg |_{SYM}$ starting from Eq. (\ref{39}) with $\Delta_0=0$. In this case
\begin{eqnarray}
n_q\bigg |_{SYM}&=&\frac{\mu+\nu+\nu_5}{\pi}-\frac{\theta(\nu-\mu-\nu_5)}{\pi}(\nu-\mu-\nu_5)%\nonumber\\&+&
+\frac{{\rm sign}(\mu-\nu_5)\theta\left (|\mu-\nu_5|-\nu\right )}{\pi}(|\mu-\nu_5|-\nu). \label{41}
\end{eqnarray}
It is easy to verify that Eqs (\ref{40}) and (\ref{41}) are identical.

\section{Phase structure}
\subsection{The role of the duality symmmetry ${\cal D}$ (\ref{16}) of the TDP}

Suppose now that at some fixed particular values of chemical potentials $\mu$, $\nu=A$ and $\nu_5=B$ the global minimum of the TDP (\ref{35}) lies at the
point, e.g., $(M=M_0\ne 0,\Delta=0)$. It means that for such fixed values
of the chemical potentials the chiral symmetry breaking (CSB) phase is
realized in the model. Then it follows from the duality invariance of the
TDP (\ref{9}) (or (\ref{35})) with respect to the transformation ${\cal D}$
(\ref{16}) that the permutation of the chemical potential values
(i.e. $\nu=B$ and $\nu_5=A$ and intact value of $\mu$) moves
the global minimum of the TDP $\Omega^{ren}(M,\Delta)$ to the point
$(M=0,\Delta=M_0)$, which corresponds to the charged PC phase
(and vice versa). This is the so-called duality correspondence
between CSB and charged PC phases in the framework of the model under
consideration. \footnote{It is worth to note that in some (1+1)- and (2+1)-dimensional models there is a duality between CSB and superconductivity \cite{thies2,ekkz2}.}

Hence, the knowledge of a phase of the model (1) at
some fixed values of external free model parameters
$\mu,\nu,\nu_5$ is sufficient  to understand what phase (we call it a dually conjugated phase) is realized at rearranged values of external parameters,
$\nu\leftrightarrow\nu_5$, at fixed $\mu$. Moreover, different physical parameters such as condensates, densities, etc, which characterize both the
initial phase and the dually conjugated phase, are connected by the duality transformation ${\cal D}$. For example, the chiral condensate of the initial CSB phase at some fixed $\mu,\nu,\nu_5$ is equal to the charged-pion condensate of the dually conjugated charged PC phase, in which one should perform the replacement  $\nu\leftrightarrow\nu_5$. Knowing the particle density $n_q(\nu,\nu_{5})$ of the initial CSB phase as a function of external chemical potentials $\nu,\nu_{5}$, one can find the particle density in the dually conjugated charged PC phase by interchanging $\nu$ and $\nu_{5}$ in the expression $n_q(\nu,\nu_{5})$ (see also in Sec. III C), etc.

The duality transformation ${\cal D}$ of the  TDP can also be
applied to an arbitrary phase portrait of the model (see below). In
particular, it is clear that if we have a most general phase
portrait, i.e. the correspondence between any point
$(\nu,\nu_5,\mu)$ of the three-dimensional space of external
parameters and possible model phases (CSB, charged PC and symmetric phase),
then under the duality transformation ($\nu\leftrightarrow\nu_5$, CSB$\leftrightarrow$charged PC) this phase
portrait is mapped to itself, i.e. the most general $(\nu,\nu_5,\mu)$-phase portrait
is self-dual. The self-duality of the $(\nu,\nu_5,\mu)$-phase portrait
means that the regions of the CSB and charged PC phases in the
three-dimensional $(\nu,\nu_5,\mu)$ space are arranged mirror
symmetrically with respect to the plane $\nu=\nu_5$ of this
space. Below, in subsection B, we will present a few sections of this three-dimensional $(\nu,\nu_5,\mu)$-phase portrait of the model by the planes of the form $\mu=const$, $\nu=const$ and $\nu_5=const$, respectively.
\begin{figure}
%----figure 1
\includegraphics[width=0.45\textwidth]{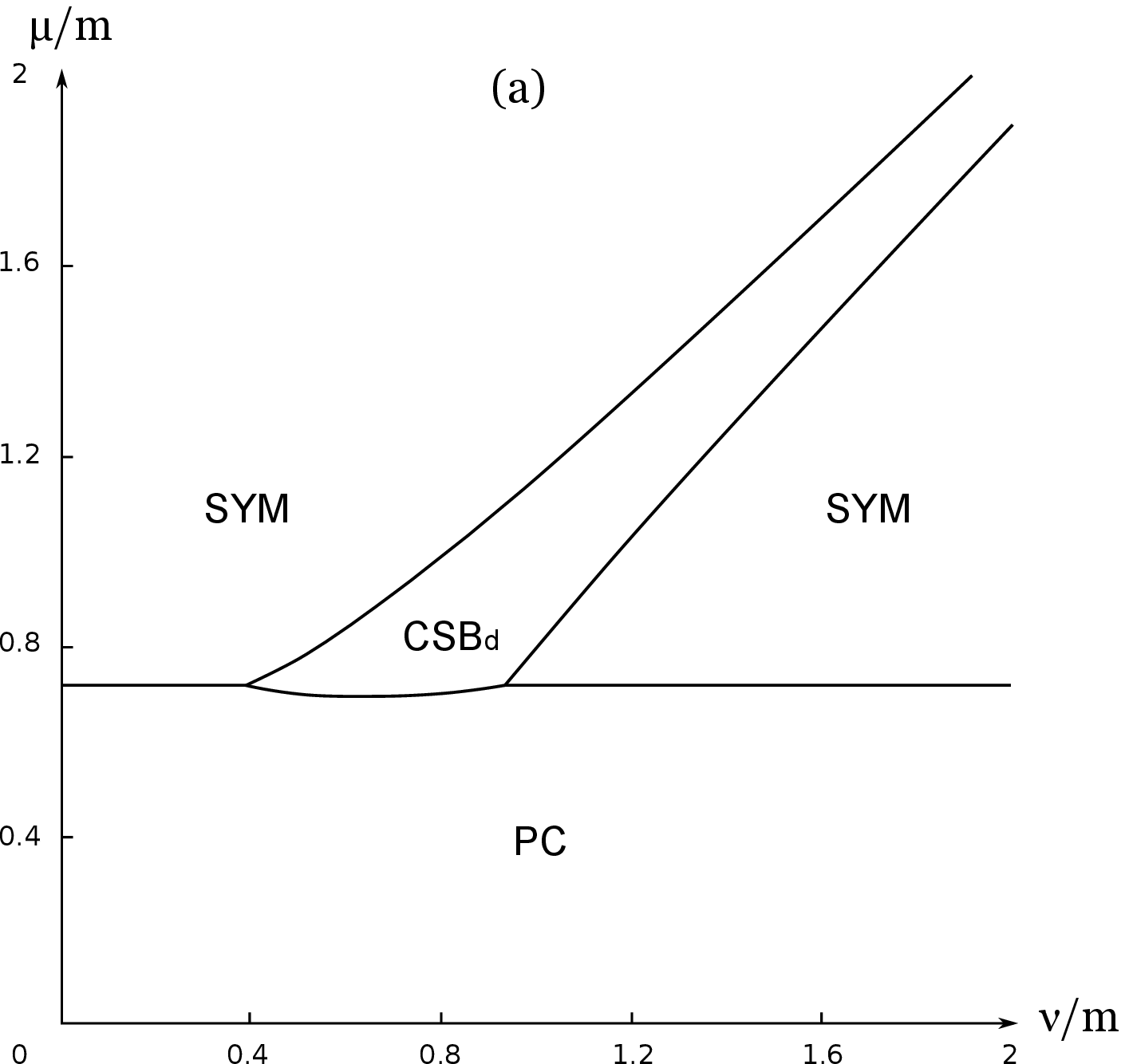}
\hfill
\includegraphics[width=0.45\textwidth]{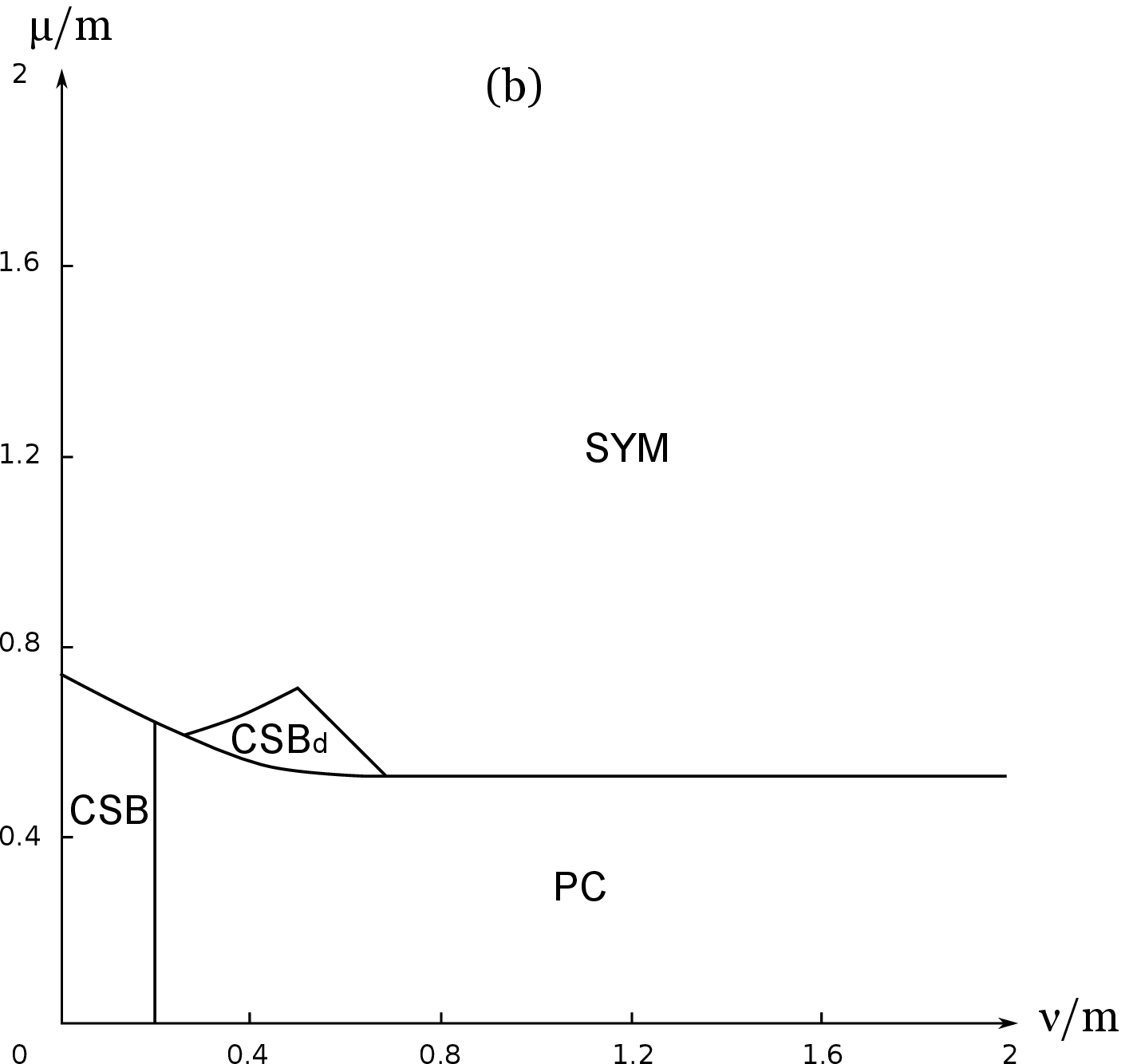}\\
\includegraphics[width=0.45\textwidth]{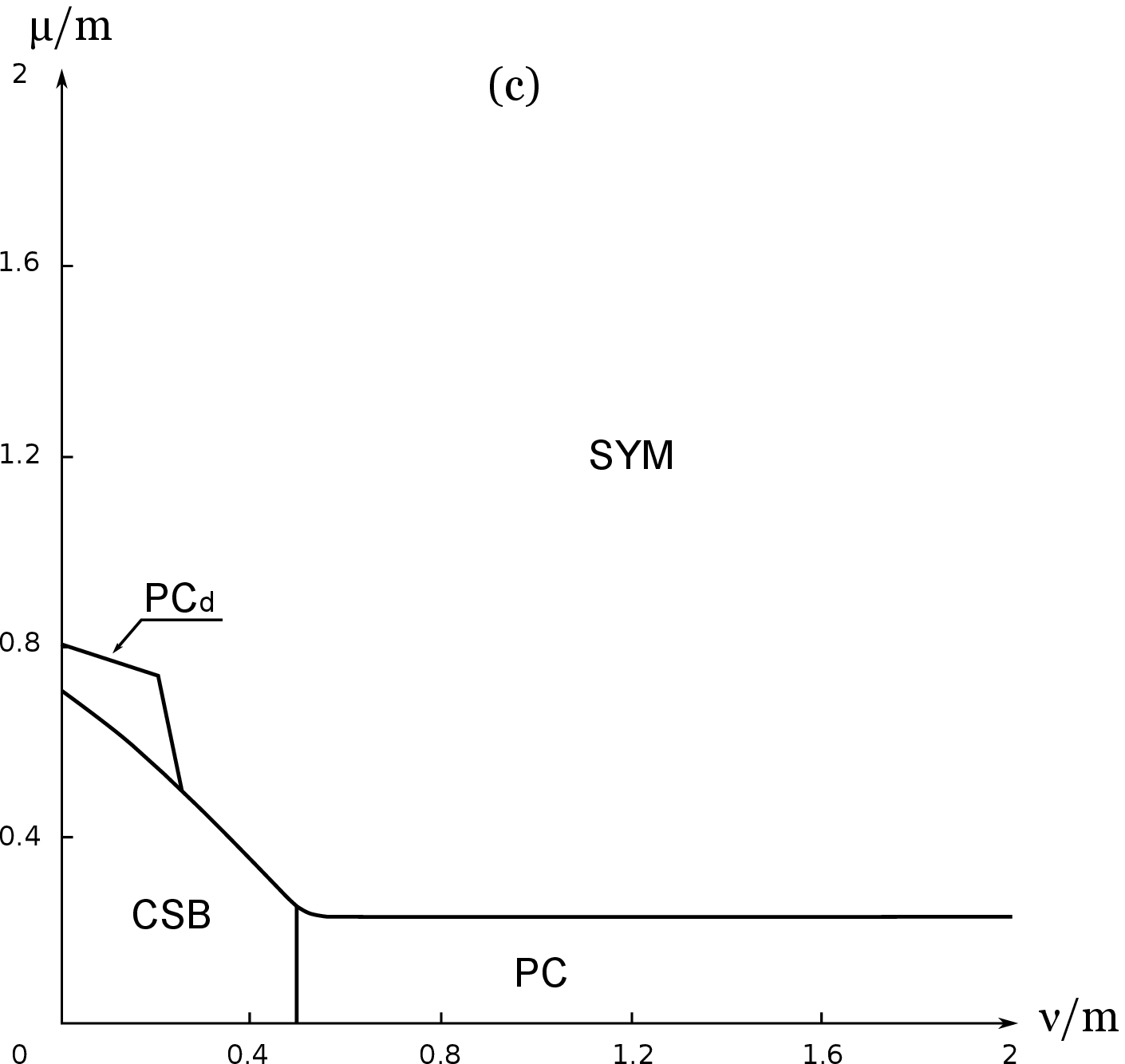}
\hfill
\includegraphics[width=0.45\textwidth]{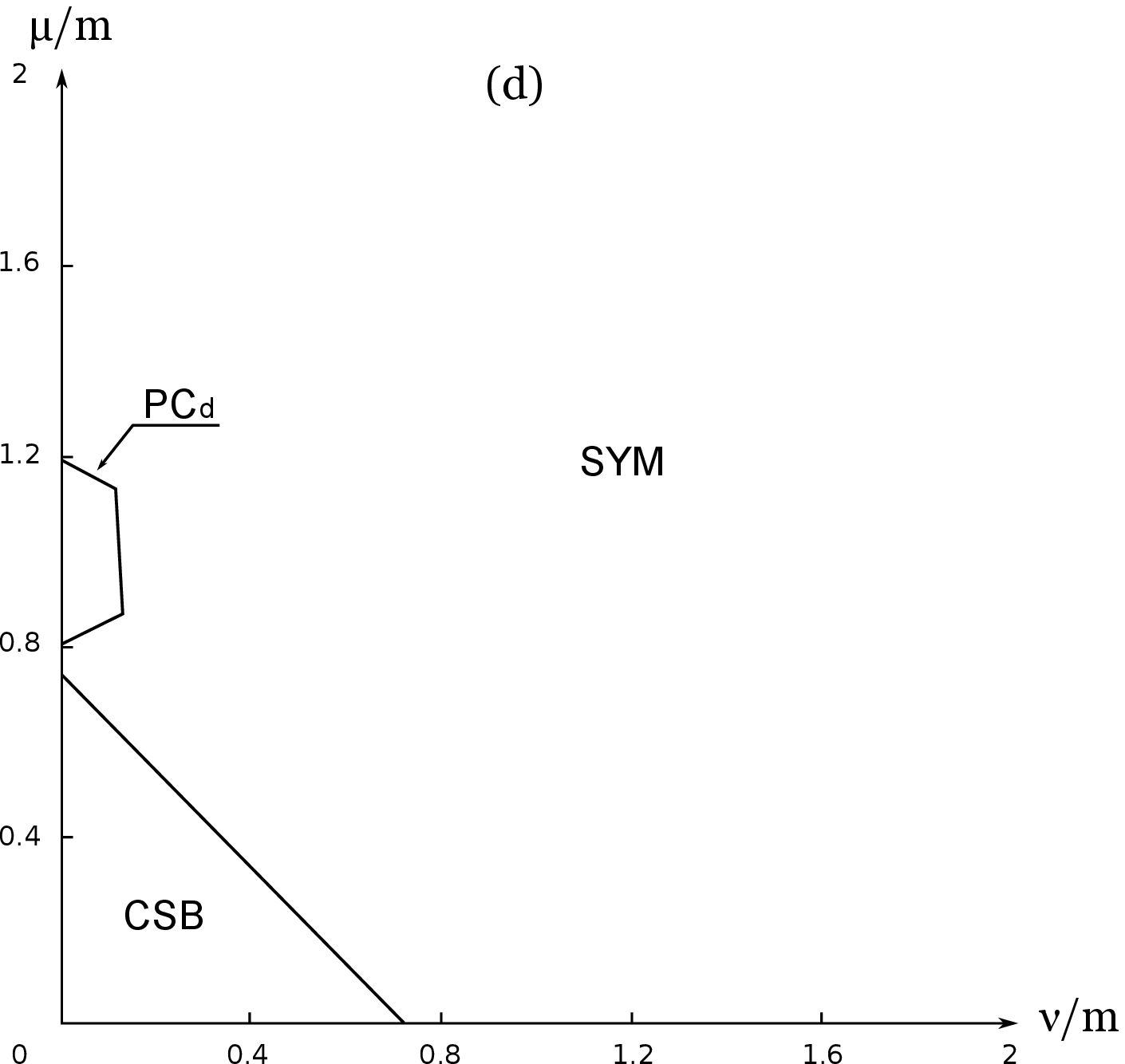}\\
%\parbox[t]{0.45\textwidth}{
\caption{The $(\nu,\mu)$-phase portrait of the model for different
values of the chiral chemical potential $\nu_5$: (a) The case
$\nu_5=0$. (b) The case  $\nu_5=0.2m$. (c) The case
$\nu_5=0.5m$. (d) The case  $\nu_5=m$. The notations PC and PCd mean
the charged pion condensation phase with zero and nonzero baryon
density, respectively. Analogously, the notations CSB and CSBd mean the chiral symmetry breaking phase with zero and nonzero baryon density, respectively,
and SYM denotes the symmetric phase. The parameter $m$ was introduced in (\ref{30}).}
\end{figure}

\subsection{Promotion of dense charged PC phase by $\nu_5\ne 0$}

First of all, we will study the phase structure of the model (1) at different fixed values of the chiral isospin chemical potential $\nu_5$. To this end, we determine numerically the global minimum points of the TDPs $F_1(M)$ (\ref{33})
and $F_2(\Delta)$ (\ref{34}) and then compare the minimum values of
these functions vs external parameters $\mu,\nu,\nu_5$. Moreover,
using the expressions (\ref{38}) and (\ref{39}), it is possible to find the
quark number density $n_q$ or baryon density $n_B$ (note that
$n_q=3n_B$) inside each phase. As a result, in Figs 1a--1d we have
drawn several $(\nu,\mu)$-phase portraits, corresponding to (a)
$\nu_5=0$, (b) $\nu_5=0.2m$, (c) $\nu_5=0.5m$, and (d)
$\nu_5=m$. Recall that $m$ is a free renormalization invariant mass scale
parameter, which appeares in the vacuum case of the model after renormalization (see 
(\ref{30}) and (\ref{31})).

The phase portrait of the model in Fig. 1a with $\nu_5=0$ was obtained earlier
(see e.g. papers \cite{ekkz,ek2}). It is clear from Fig. 1a that at
$\nu_5=0$ the charged PC phase with {\it nonzero baryon density} $n_B$
(in Figs 1b--1d it is denoted by the symbol PCd) is not realized in the model under consideration. Only the charged PC phase with {\it zero baryon density} can be observed at rather small values of $\mu$. (Physically, it means that at $\nu_5=0$ the model predicts the charged PC phenomenon in the medium with $n_B=0$ only. For example, it might consist of charged pions, etc. But in quark matter with nonzero baryon density the charged PC is forbidden.) Instead, at large values of $\mu$ there exist two phases, the chiral symmetry breaking and the symmetrical one, both with nonzero baryon density, i.e. the model predicts the CSB phase of dense quark matter. However, as we can see from other phase diagrams of Fig. 1, at rather high values of $\nu_5$ there might appear on the phase portrait a charged PC phase with {\it nonzero baryon density} (it is denoted as PCd in Figs 1). Hence, in chirally asymmetric, i.e. for $\nu_5>0$, and dense quark matter the charged PC phenomenon
is allowed to exist in the framework of the toy model (1). Thus, we see that $\nu_5\ne 0$ is a factor which promotes the charged PC phenomenon in dense quark matter. Note that the compact region of the $(\nu,\mu)$-plane, which is occupied by the PCd phase (see, e.g., in Fig. 1d at $\nu_5=m$), continues to move up along the $\mu$-axis, when $\nu_5$ increases above the value $\nu_5=m$.

Now, suppose that we want to obtain a $(\nu_5,\mu)$-phase portrait of
the model at some fixed value $\nu=const$. In this case there is no
need to perform the direct numerical investigations of the TDP
(\ref{35}). In contrast (due to the dual invariance (\ref{16}) of the
model TDP), one can simply make the dual transformation of the
$(\nu,\mu)$-phase diagram at the corresponding fixed value
$\nu_5=const$. For example, to find the $(\nu_5,\mu)$-phase diagram at
$\nu=0$ we should start from the $(\nu,\mu)$-diagram at fixed
$\nu_5=0$ of Fig. 1a and make the simplest replacement in the
notations of this figure: $\nu\to\nu_5$, PC$\leftrightarrow$CSB,
PCd$\leftrightarrow$CSBd. As a result of this mapping, we obtain the
phase diagram of Fig. 2a with PCd phase. In a similar way, to obtain the $(\nu_5,\mu)$-phase diagram at $\nu=0.2m$, it is sufficient to apply the duality transformation to Fig. 1b (recall, it is the $(\nu,\mu)$-phase portrait of the model at $\nu_5=0.2m$). The resulting mapping is Fig. 2b, etc. It thus supports the above conclusion: the charged PC phenomenon can be realized in chirally asymmetric quark matter with nonzero baryon density.
\begin{figure}
%----figure 2
\includegraphics[width=0.45\textwidth]{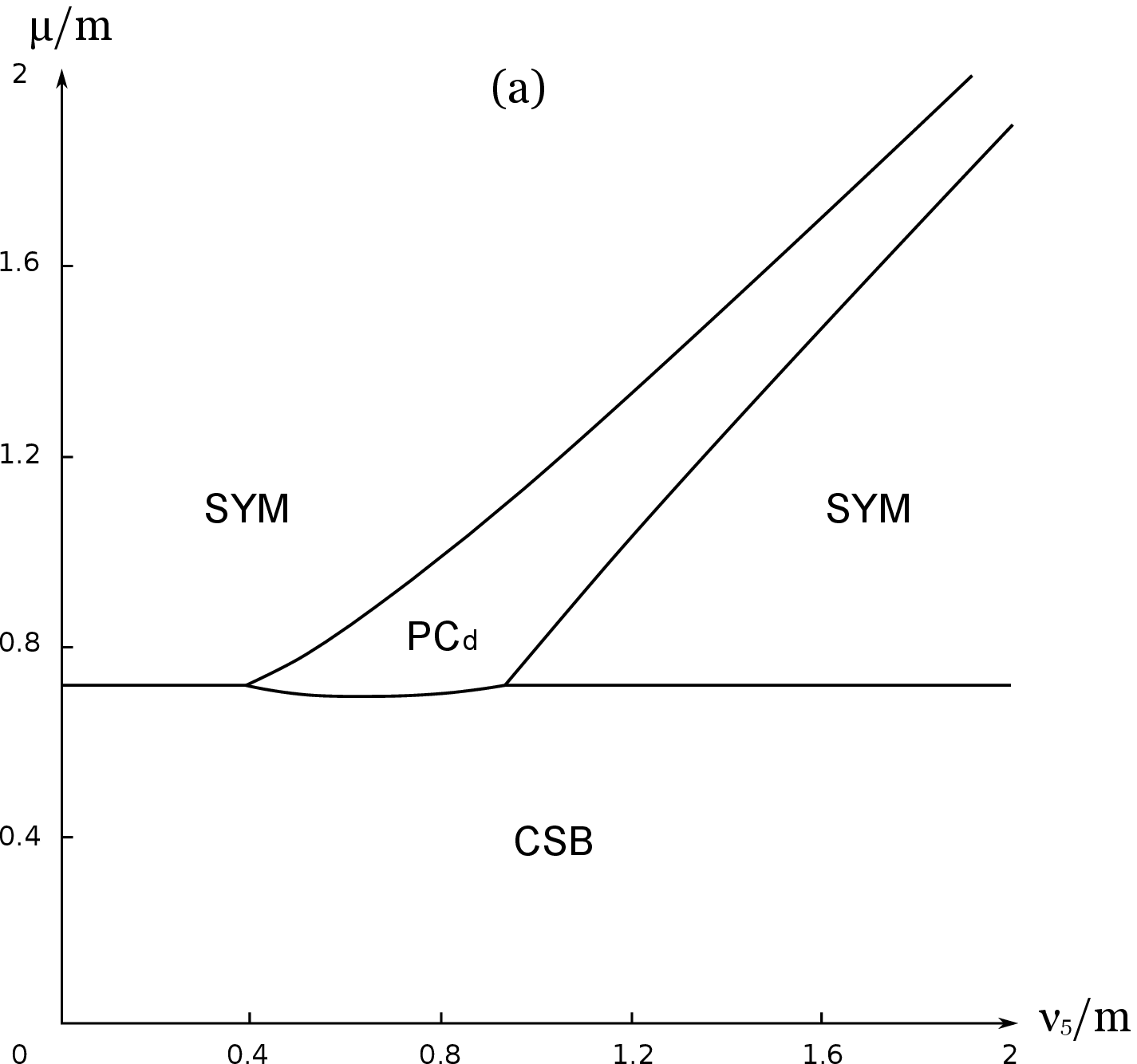}
\hfill
\includegraphics[width=0.45\textwidth]{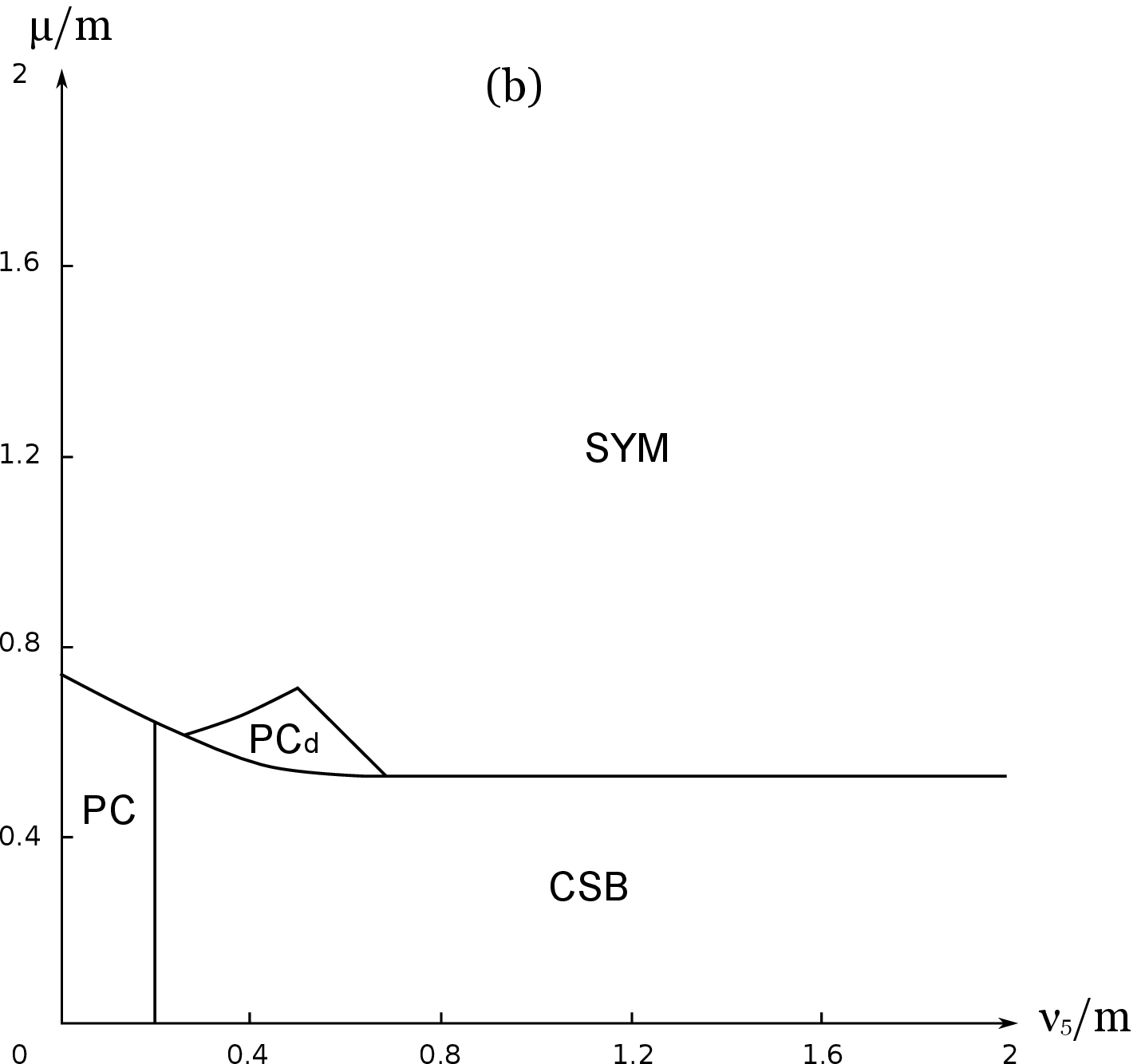}
%\parbox[t]{0.45\textwidth}{
\caption{The $(\nu_5,\mu)$-phase portrait of the model for different
values of the isospin chemical potential $\nu$: (a) The case
$\nu=0$. (b) The case  $\nu=0.2m$. Other notations are the same as in Fig. 1.}
\end{figure}

Finally, let us consider the $(\nu,\nu_5)$-phase diagrams of the model
at different fixed values of $\mu$. It is clear from the previous
discussions that each of these diagrams is a self-dual one, i.e. the CSB and charged PC phases are arranged symmetrically with respect to the line $\nu=\nu_5$ of the $(\nu,\nu_5)$-plane. This fact is confirmed by several $(\nu,\nu_5)$-phase portraits in Fig. 3, obtained by direct numerical analysis of the TDPs $F_1(M)$ (\ref{33}) and $F_2(\Delta)$ (\ref{34}). Moreover, the phase diagrams of Fig. 3 support once again the main conclusion of our paper: the charged PC phase with nonzero baryon density, i.e. the phase denoted in Figs 1--3 as PCd, might be realized in the framework of the model (1) only at $\nu_5>0$.
\begin{figure}
%----figure 3
\includegraphics[width=0.45\textwidth]{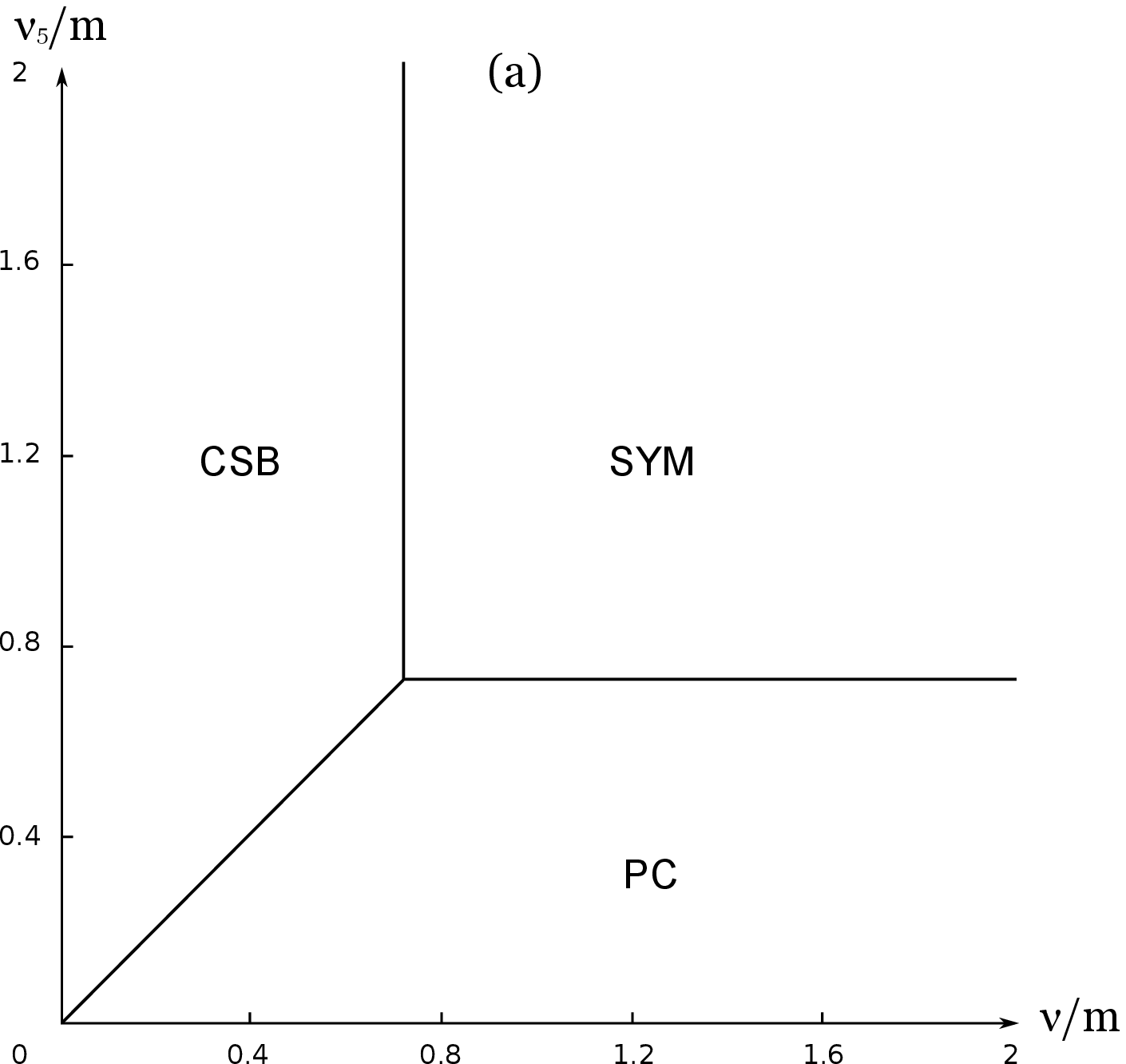}
\hfill
\includegraphics[width=0.45\textwidth]{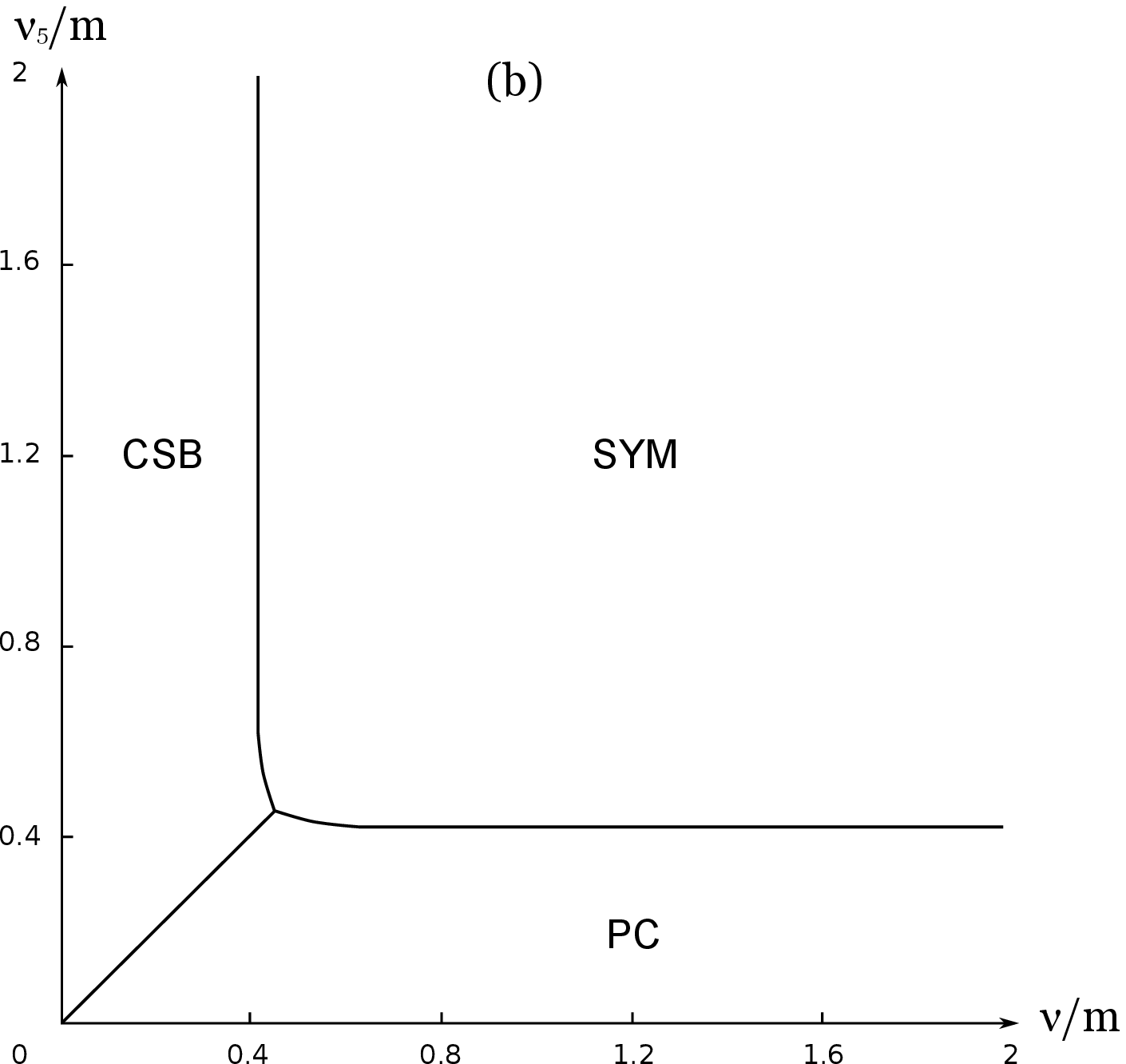}\\
\includegraphics[width=0.45\textwidth]{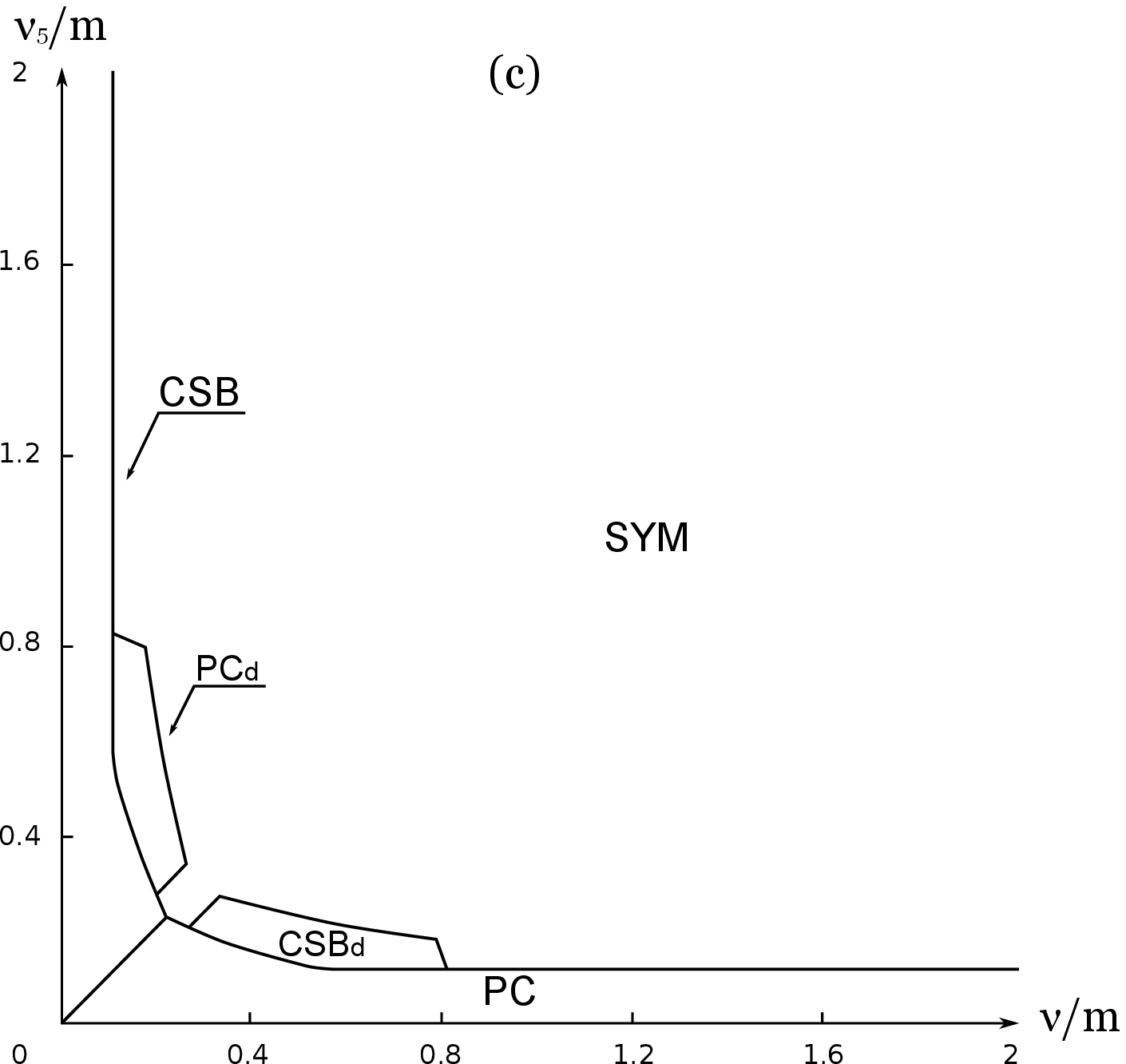}
\hfill
\includegraphics[width=0.45\textwidth]{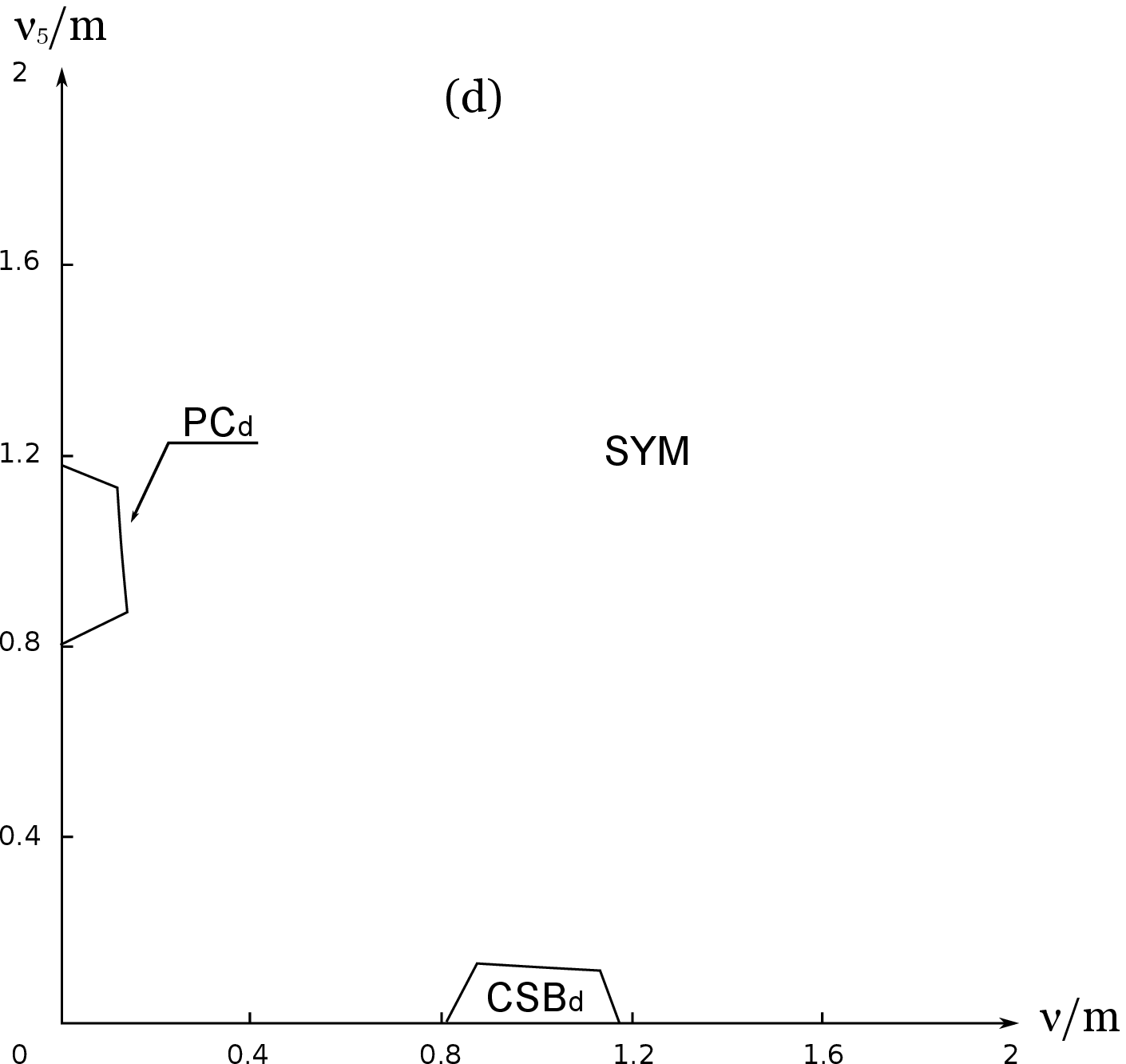}\\
%\parbox[t]{0.45\textwidth}{
\caption{The $(\nu,\nu_5)$-phase portrait of the model for different
values of the  quark number chemical potential $\mu$: (a) The case
$\mu=0$. (b) The case  $\mu=0.3m$. (c) The case  $\mu=0.6m$. (d) The
case  $\mu=m$. Other notations are the same as in Fig. 1.}
\end{figure}
\begin{figure}
%----figure 4
\includegraphics[width=0.45\textwidth]{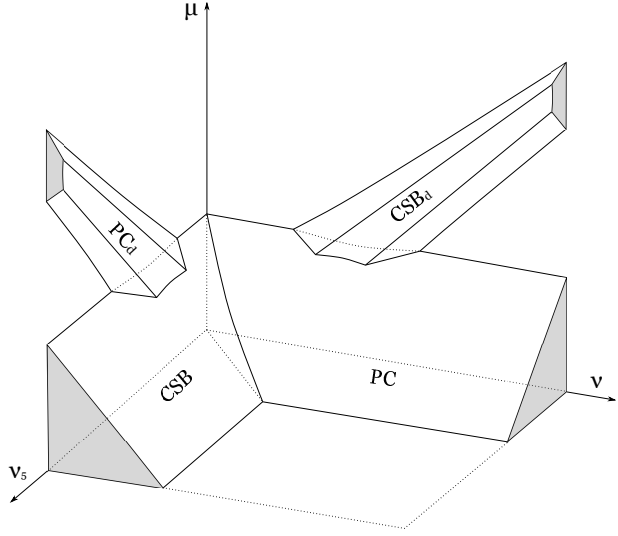}
\caption{Schematic representation of the model phase portrait in the $(\nu,\nu_5,\mu)$-parameter space. The notations are the same as in Fig. 1. The points which are outside PC-, CSB-, PCd-, and CSBd phases of the diagram correspond to the symmetric phase.}
\end{figure}

Taking into account the particular phase diagrams of Figs 1--3, it is possible to represent schematically the most general phase portrait of the model in the space of chemical potentials $\nu,\nu_5,\mu$ (see Fig.4).

\section{Summary and conclusions}

In this paper, the phase structure of the NJL$_2$ model (1) with two quark flavors is investigated in the large-$N_c$ limit in the presence of baryon
$\mu_B$, isospin $\mu_I$ and chiral isospin $\mu_{I5}$ chemical
potentials. For the particular case with $\mu_{I5}=0$, the task was
solved earlier in Refs \cite{ekkz,massive,ek2}, where it was shown
that the toy model (1) does not predict a charged PC phase of dense and
isotopically asymmetric quark matter. So our present consideration is
a generalization of this approach to the case $\mu_{I5}\ne 0$, i.e. it
is devoted, although in the framework of a simpler (1+1)-dimensional
model, to the study of the properties of chirally ($\mu_{I5}\ne 0$)
and isotopically ($\mu_{I}\ne 0$) asymmetric  dense ($\mu_{B}\ne 0$)
quark matter. The following two new physical effects are predicted:

1) It is clear from the phase diagrams of Figs 1--3 that the charged
PC phase with nonzero baryon density (this phase is denoted in Figs
1--3 by the symbol PCd), prohibited at $\mu_{I5}=0$, might appear at rather large values of $\mu_{I5}>0$. Hence, chiral asymmerty (i.e. $\mu_{I5}\ne 0$ in (1))
of dense quark matter can serve as a factor promoting there a charged
pion condensation phenomenon. Note that two other known possibilities
to generate a charged PC phase of dense quark matter in model (1)
are: (i) to put a system into a finite volume \cite{ekkz} or (ii) to
take into account the possibility for a spatial inhomogeneity of condensates \cite{gkkz}. 

2) We have shown in the leading order of the large-$N_c$ approximation
that in the framework of the NJL$_2$ model (1) there is a duality
correspondence between CSB and charged PC phenomena. It means that if, e.g.,
for some initial fixed set of external parameters $(\mu_B,\mu_I=A,\mu_{I5}=B)$, 
the chiral symmetry breaking phase is realized in the model, then for
a rearranged set of external parameters, i.e. for the set
$(\mu_B,\mu_I=B,\mu_{I5}=A)$, the so-called dually conjugated charged
PC phase is arranged (and vice versa). It must be emphasized that
different physical quantities such as order parameter (condensate), particle density, etc of the initial phase and its dually conjugated one are equal. In this way, it is sufficient to have the information about the ground state of the initial phase, which is realized for the set $(\mu_B,\mu_I,\mu_{I5})$, in order to determine
the properties of the ground state of the dually conjugated phase,
corresponding to the rearranged external parameter set $(\mu_B,\mu_{I5},\mu_I)$.
(Recall that another kind of duality, the duality between CSB and superconductivity, exists also in some (1+1)- and (2+1)-dimensional NJL models \cite{thies2,ekkz2}.)

It was shown recently that in the large-$N_c$ limit there is an equivalence (duality) between the phase structure of the  $SU(N_c)$ QCD at finite $\mu_I$ and the phase structures of some QCD-like models at finite $\mu_B$. Moreover, if $\mu_I$ is outside the BEC-BCS crossover region, then there might exist an equivalence between chiral symmetry breaking and charged pion condensation within the QCD itself (see, e.g., Ref. \cite{Hanada:2011ju} and, in particular, Fig. 1 there). In a similar way it was shown that QCD at $\mu_{I5}\ne 0$ is equivalent to QCD at $\mu_{I}\ne 0$ in the chiral limit and at $N_c\to\infty$ (see Sec. 4 in   \cite{Hanada:2011jb}). These facts are the basis for some hope that the present general analysis of the (1+1)-dimensional toy model (1) with three nonzero chemical potentials $\mu$, $\mu_{I}$ and $\mu_{I5}$ will shed some new light on physical effects in chirally and isotopically asymmetric dense quark matter in the real (3+1)-dimensional QCD at large $N_c$. Furthermore, we believe that at large $N_c$ there is (in the chiral limit) a duality between chiral symmetry breaking and charged pion condensation in the (3+1)-dimensional two flavor  NJL model in the presence of the isospin and chiral isospin chemical potentials. The check of this assumption is our next goal. 
  
\appendix
\section{Evaluation of the roots of the polynomial $P_4(p_0)$ (\ref{91}) }
\label{ApB}

\subsection{General case}
\vspace{-0.4cm}

It is very convenient to present the fourth-order polynomial (\ref{91}) of the variable $\eta\equiv p_0+\mu$  as a product of two second-order polynomials (this way is proposed in \cite{Birkhoff}), i.e. we assume that
\begin{eqnarray}
\eta^4&-&2a \eta^2-b \eta +c = (\eta^2 + r\eta + q)(\eta^2 - r\eta + s)\nonumber\\
&&=\left [\left (\eta+\frac r2\right )^2+q-\frac{r^2}{4}\right ]\left [\left
(\eta-\frac r2\right )^2+s-\frac{r^2}{4}\right ]\equiv
(\eta-\eta_{1})(\eta-\eta_{2})(\eta-\eta_{3})(\eta-\eta_{4}),\label{B3}
\end{eqnarray}
where $r$, $q$ and $s$ are some real valued quantities, such that
(see the relations (\ref{10})):
\begin{eqnarray}
-2a~&& \equiv-2(M^2+\Delta^2+p_1^2+\nu^2+\nu_{5}^2)= s+q-r^2;~~ -b\equiv -8p_1\nu\nu_{5}= rs-qr;\nonumber\\
c~&&\equiv a^2-4p_1^2(\nu^2+\nu_5^2)-4M^2\nu^2-4\Delta^2\nu_5^2-4\nu^2\nu_5^2=sq.
\label{B4}
\end{eqnarray}
In the most general case, i.e. at $M\ge 0$, $\Delta\ge 0$, $\nu\ge 0$,$\nu_5\ge 0$ and arbitrary values of $p_1$, one can solve the system of equations (\ref{B4}) with respect to $q,s,r$ and find
\begin{eqnarray}
q=\frac 12 \left (-2a +R+\frac{b}{\sqrt{R}}\right ),~~ s=\frac 12 \left
(-2a +R-\frac{b}{\sqrt{R}}\right ),~~r=\sqrt{R},\label{B5}
\end{eqnarray}
where $R$ is an arbitrary real positive solution of the equation
\begin{eqnarray}
\label{B6} X^3 + AX=BX^2 + C
\end{eqnarray}
with respect to a variable $X$, and
\begin{eqnarray}
A&=& 4a^2-4c=16\Big[\nu_5^2\Delta^2+M^2\nu^2+
\nu_5^2\nu^2+p_1^2(\nu^2+\nu_5^2)\Big],\nonumber\\
B&=&4a =4(M^2+\Delta^2+\nu^2+\nu_5^2+p_1^2),~~ C=b^2=(8\nu_5\nu
p_1)^2. \label{B7}
\end{eqnarray}
Finding (numerically) the quantities $q$, $s$ and $r$, it is possible to obtain from (\ref{B3}) the roots $\eta_i$:
\begin{eqnarray}
\eta_{1}=-\frac{r}{2}+\sqrt{\frac{r^2}{4}-q},~~\eta_{2}=\frac{r}{2}+
\sqrt{\frac{r^2}{4}-s},~~\eta_{3}=-\frac{r}{2}-\sqrt{\frac{r^2}{4}-q},~~\eta_{4}=\frac{r}{2}-\sqrt{\frac{r^2}{4}-s}.\label{B41}
\end{eqnarray}
Numerical investigation shows that in the most general case the discriminant of the third-order algebraic equation (\ref{B6}), i.e. the quantity $18ABC-4B^3C+A^2B^2-4A^3-27C^2$, is always nonnegative. So the equation (\ref{B6}) vs $X$ has three real solutions $R_1,R_2$ and $R_3$ (this fact is presented in \cite{Birkhoff}). Moreover, since the coefficients $A$, $B$ and $C$ (\ref{B7})
are nonnegative, it is clear that, due to the form of equation (\ref{B6}), all its roots $R_1$, $R_2$ and $R_3$ are also nonnegative quantities (usually, they are positive and different). So we are free to choose the quantity $R$ from (\ref{B5}) as one of the positive solutions $R_1$, $R_2$ or $R_3$. In each case, i.e. for
$R=R_1$, $R=R_2$, or $R=R_3$, we will obtain the same set of 
roots (\ref{B41}) (possibly rearranged), which depends only on $\nu$, $\nu_5$, $M$, $\Delta$ and $p_1$, and does not depend on the choice of $R$. Due to the relations (\ref{B3})-(\ref{B41}), one can find numerically (at fixed values of  $\mu$, $\nu$, $\nu_5$, $M$, $\Delta$ and $p_1$) the roots $\eta_i=p_{0i}+\mu$ (\ref{B41}) and, as a result, investigate numerically the TDP (\ref{28}). It is clear also from
(\ref{B3})-(\ref{B41}) that the roots $\eta_i$ are even functions vs
$p_1$. So in all improper $p_1$ integrals, which include quasiparticle
energies $p_{0i}$ (see, e.g., the integral in Eq. (\ref{28})), we can restrict
ourselves to an integration over nonnegative values of $p_1$ (up to a factor 2).

On the basis of the relations (\ref{B3})-(\ref{B41}) let us consider the asymptotic behavior of the quasiparticle energies $p_{0i}$ at $p_1\to\infty$. First of all, we start from the asymptotic analysis of the roots $R_{1,2,3}$ of the equation (\ref{B6}) at $p_1\to\infty$,
\begin{eqnarray}
\label{B701}
R_1&=&4\nu^2-\frac{4\Delta^2\nu^2}{p_1^2}
+{\cal O}\big (1/p_1^4\big ),\\
\label{B7001}
R_2&=&4\nu_5^2-\frac{4M^2\nu_5^2}{p_1^2}
+{\cal O}\big (1/p_1^4\big ),\\
R_3&=&4p_1^2+4(M^2+\Delta^2)+\frac{4(\nu_5^2M^2+\nu^2\Delta^2)}{p_1^2}
+{\cal O}\big (1/p_1^4\big ). \label{B71}
\end{eqnarray}
It is clear from these relations that $R_3$ is invariant under the duality transformation (\ref{16}), whereas $R_1\leftrightarrow R_2$.
Then, using for example $R_3$ (\ref{B71}) as the quantity $R$ in Eqs. (\ref{B5}) and (\ref{B41}), one can get the asymptotics of the quasiparticle energies $p_{0i}\equiv \eta_i-\mu$ at $p_1\to\infty$,
\begin{eqnarray}
p_{01}&=&-|p_1|-\mu+|\nu_5-\nu|-\frac{\Delta^2+M^2}{2|p_1|} +{\cal O}\big
(1/p_1^2\big ),~~ p_{02}=|p_1|-\mu+\nu_5+\nu+\frac{\Delta^2+M^2}{2|p_1|}
+{\cal O}\big (1/p_1^2\big ),\nonumber\\
p_{03}&=&-|p_1|-\mu-|\nu_5-\nu|-\frac{\Delta^2+M^2}{2|p_1|} +{\cal O}\big
(1/p_1^2\big ),~~ p_{04}=|p_1|-\mu-\nu_5-\nu+\frac{\Delta^2+M^2}{2|p_1|}
+{\cal O}\big (1/p_1^2\big ).\label{B26}
\end{eqnarray}
Finally, it follows from (\ref{B26}) that at $p_1\to\infty$
\begin{eqnarray}
|p_{01}|+|p_{02}|+|p_{03}|+|p_{04}|=4|p_1|+\frac{2(\Delta^2+M^2)}{|p_1|}
+{\cal O}\big (1/p_1^2\big ).\label{B9}
\end{eqnarray}
For the purposes of the renormalization of the TDP (\ref{28}), it is
very important that the leading terms of this asymptotic behavior do
not depend on different chemical potentials, i.e. the quantity
$\sum_{i=1}^4|p_{0i}|$ at $\mu=\nu=\nu_5=0$ has the same asymptotics
(\ref{B9}).  Moreover, we would like to emphasize once again that the
asymptotic behavior (\ref{B9}) does not depend on which of the roots
$R_1$, $R_2$ or $R_3$ of the equation (\ref{B6}) is taken as the
quantity $R$ in the relations (\ref{B5}).\vspace{-0.6cm}

\subsection{Consideration of some particular cases}
\vspace{-0.4cm}

Note that in some particular cases it is possible to solve exactly the  third order auxiliary equation (\ref{B3}) and, as a result, to present the  quasiparticle energies $p_{0i}$ (or the roots $\eta_i$ of the polynomial (\ref{B3})) in an explicit analytical form.

{\bf\underline{1. The case $\mu=\nu=\nu_5=0$}}.
It is clear from (\ref{B6}) and (\ref{B7}) that at $\nu=\nu_5=0$ we have $A=C=0$, so $R_{1,2}=0$, $R_3=4(M^2+\Delta^2+p_1^2)$. In this case $q=s=r^2/4=M^2+\Delta^2+p_1^2$ and $\eta_{1,2}=\sqrt{M^2+\Delta^2+p_1^2}$, $\eta_{3,4}=-\sqrt{M^2+\Delta^2+p_1^2}$. If in addition $\mu=0$, then we have
\begin{eqnarray}
\big (|p_{01}|+|p_{02}|+|p_{03}|+|p_{04}|\big )\Big |_{\mu=\nu=\nu_5=0}=
4\sqrt{M^2+\Delta^2+p_1^2}.\label{B10}
\end{eqnarray}
As was noted above, this quantity at $p_1\to\infty$ is expanded in the form  (\ref{B9}).

{\bf\underline{2. The case $\Delta=0$}}. In this particular case the exact expression for the set of quasiparticle energies $p_{0i}$ was already presented in (\ref{26}). Here we would like to demonstrate how this result is reproduced in the framework of the procedure (\ref{B3})-(\ref{B41}).

It is easy to see that at $\Delta=0$ there is an evident root $R_1=4\nu^2$ of the polynomial (\ref{B6}). On this basis we can find exact expressions for the other two its roots,
\begin{eqnarray}
R_{2,3}=2(M^2+\nu_5^2+p_1^2)\pm 2\sqrt{(M^2+\nu_5^2+p_1^2)^2-4\nu_5^2p_1^2}=(E_1\pm E_2)^2,\label{B11}
\end{eqnarray}
where
\begin{eqnarray}
E_1=\sqrt{M^2+(p_1+\nu_5)^2},~~~E_2=\sqrt{M^2+(p_1-\nu_5)^2}.\label{B12}
\end{eqnarray}

If $R_1=4\nu^2$ is taken as the quantity $R$ of the relations (\ref{B5}), then, using (\ref{B5}) in (\ref{B41}), we obtain directly the expression (\ref{26}) for the set of quasiparticle energies $p_{0i}$.

If, e.g., $R=R_3\equiv (E_1+E_2)^2$, then, taking into account the evident relation $E_1^2-E_2^2=4p_1\nu_5$, we have from (\ref{B5})
\begin{eqnarray}
r=E_1+E_2,~~~q=E_1E_2-\nu^2&+&\nu (E_1-E_2),~~~s=E_1E_2-\nu^2-\nu (E_1-E_2),\nonumber\\
\frac{r^2}{4}-q=\frac{(E_1-E_2-2\nu)^2}{4},&~&\frac{r^2}{4}-s=\frac{(E_1-E_2+2\nu)^2}{4}.\label{B13}
\end{eqnarray}
Using these relations in (\ref{B41}), we receive for the quasiparticle energies $p_{0i}$ the same set as in (\ref{26}). Thereby we have demonstrated that the set of roots $\eta_i$ (\ref{B41}) does not depend on which of the solutions $R_1$, $R_2$ or $R_3$ of the equation (\ref{B6}) is used as the quantity $R$ in the relations (\ref{B5}).

{\bf\underline{3. The case $M=0$}}. In a similar way it is possible to show that Eq. (\ref{B6}) at $M=0$ has the following three roots:
\begin{eqnarray}
R_1=4\nu_5^2,~~R_{2,3}=({\cal E}_1\pm {\cal E}_2)^2,\label{B14}
\end{eqnarray}
where
\begin{eqnarray}
{\cal E}_1=\sqrt{\Delta^2+(p_1+\nu)^2},~~~{\cal E}_2=\sqrt{\Delta^2+(p_1-\nu)^2}.\label{B15}
\end{eqnarray}
On the basis of each of them, using the relations (\ref{B41}) and (\ref{B5}), one can obtain the set of quasiparticle energies (\ref{27}).

{\bf\underline{4. The case $\nu_5=\nu$}}. In this particular case Eq. (\ref{B6}) has the following three roots:
\begin{eqnarray}
R_1=4\nu^2,~~R_{2,3}=(\widetilde E_1\pm \widetilde E_2)^2,\label{B16}
\end{eqnarray}
where
\begin{eqnarray}
\widetilde E_1=\sqrt{M^2+\Delta^2+(p_1+\nu)^2},~~~\widetilde E_2=\sqrt{M^2+\Delta^2+(p_1-\nu)^2}.\label{B17}
\end{eqnarray}
Taking for simplicity $R=R_1$ in (\ref{B5}) and using the relations (\ref{B41}), we have in this case for the quasiparticle energies $p_{0i}$ the following set of values:
\begin{eqnarray}
\Big\{p_{01},p_{02},p_{03},p_{04}\Big\}\Big |_{\nu_5=\nu}=\Big\{-\mu-\nu\pm\sqrt{M^2+\Delta^2+(p_1-\nu)^2},-\mu+\nu\pm\sqrt{M^2+\Delta^2+(p_1+\nu)^2}\Big\}.
\end{eqnarray}

\section{Derivation of the relation (\ref{33})}
\label{ApD}

If $\Delta=0$ and $M\ne 0$, then the quasiparticle energies $p_{0i}$ are presented in the expression (\ref{26}). So
\begin{eqnarray}
\big (|p_{01}|+|p_{02}|+|p_{03}|+|p_{04}|\big )\big |_{\Delta=0}&=&\nonumber\\
\sum_{\kappa=\pm}\Big (\left
|-\mu+\kappa\nu+\sqrt{M^2+(p_1+\kappa\nu_5)^2}\right |&&\hspace{-4mm}+\left
|-\mu+\kappa\nu-\sqrt{M^2+(p_1+\kappa\mu_5)^2}\right |\Big )\nonumber\\
=2\sum_{\kappa=\pm}\Bigg\{\sqrt{M^2+(p_1+\kappa\nu_5)^2}&&\hspace{-4mm}+
\Big (\mu-\kappa\nu-\sqrt{M^2+(p_1+\kappa\nu_5)^2}\Big )
\theta \Big (\mu-\kappa\nu-\sqrt{M^2+(p_1+\kappa\nu_5)^2}\Big )\nonumber\\
+\Big (\kappa\nu-\mu&&\hspace{-4mm}-\sqrt{M^2+(p_1+\kappa\nu_5)^2}\Big )
\theta \Big (\kappa\nu-\mu-\sqrt{M^2+(p_1+\kappa\nu_5)^2}\Big )\Bigg\}, \label{36}
\end{eqnarray}
where we have took into account the well-known relations $|x|=x\theta
(x)-x\theta (-x)$ and $\theta (x)=1-\theta(-x)$. Hence, the expression (\ref{35}) at $\Delta=0$ and $M\ne 0$ can be presented in the following form:
\begin{eqnarray}
\Omega^{ren} (M,\Delta=0)&=&-\frac{M^2}{2\pi}+\frac{M^2}{2\pi}\ln\left
(\frac{M^2}{m^2}\right )-U-V, \label{D1}
\end{eqnarray}
where
\begin{eqnarray}
U&=&\int_0^\infty\frac{dp_1}{\pi}\Big\{\sqrt{M^2+(p_1+\nu_5)^2}+\sqrt{M^2+(p_1-\nu_5)^2}-2\sqrt{M^2+p_1^2}\Big\}=\frac{\nu_5^2}{\pi},\label{D2}\\
V&=&\sum_{\kappa=\pm}\int_0^\infty\frac{dp_1}{\pi}\Big\{\Big (\mu-\kappa\nu-\sqrt{M^2+(p_1+\kappa\nu_5)^2}\Big
) \theta\Big (\mu-\kappa\nu-\sqrt{M^2+(p_1+\kappa\nu_5)^2}\Big )\nonumber\\&&~~~~~~~+\Big
(\kappa\nu-\mu-\sqrt{M^2+(p_1+\kappa\nu_5)^2}\Big ) \theta\Big
(\kappa\nu-\mu-\sqrt{M^2+(p_1+\kappa\nu_5)^2}\Big )\Big\}\label{D20}\\
&=&\int_0^\infty\frac{dp_1}{\pi}\Big (\mu-\nu-\sqrt{M^2+(p_1+\nu_5)^2}\Big ) \theta\Big (\mu-\nu-\sqrt{M^2+(p_1+\nu_5)^2}\Big )\nonumber\\&+&\int_0^\infty\frac{dp_1}{\pi}\Big
(\nu-\mu-\sqrt{M^2+(p_1+\nu_5)^2}\Big ) \theta\Big
(\nu-\mu-\sqrt{M^2+(p_1+\nu_5)^2}\Big )\nonumber\\
&+&\int_0^\infty\frac{dp_1}{\pi}\Big (\mu+\nu-\sqrt{M^2+(p_1-\nu_5)^2}\Big
) \theta\Big (\mu+\nu-\sqrt{M^2+(p_1-\nu_5)^2}\Big ).\label{D3}
\end{eqnarray}
Notice that a calculation of the convergent improper integral $U$
(\ref{D2}) can be found, e.g., in Appendix C of \cite{ekkz}. Moreover,
when summing in (\ref{D20}) over $\kappa =\pm$, we took into account
that $\mu\ge 0$ and $\nu\ge 0$. So there are only three integrals in
the expression (\ref{D3}). Due to the presence of the step function
$\theta (x)$, each integral in (\ref{D3}) is indeed a proper one. Let
us denote the sum of the first two integrals of (\ref{D3}) as $V_1$
and the last integral as $V_2$, i.e. $V=V_1+V_2$. Then, it is evident that
\begin{eqnarray}
V_1&=&\int_0^\infty\frac{dp_1}{\pi}\Big (|\mu-\nu|-\sqrt{M^2+(p_1+\nu_5)^2}\Big ) \theta\Big (|\mu-\nu|-\sqrt{M^2+(p_1+\nu_5)^2}\Big ),\label{D4}\\
V_2&=&\int_0^\infty\frac{dp_1}{\pi}\Big (\mu+\nu-\sqrt{M^2+(p_1-\nu_5)^2}\Big ) \theta\Big (\mu+\nu-\sqrt{M^2+(p_1-\nu_5)^2}\Big ).\label{D5}
\end{eqnarray}
Carring out in the integrals (\ref{D4}) and (\ref{D5}) the change of variables, $q=p_1+\nu_5$ and  $q=p_1-\nu_5$, respectively, we have
\begin{eqnarray}
V_1&=&\int_{\nu_5}^\infty\frac{dq}{\pi}\Big (|\mu-\nu|-\sqrt{M^2+q^2}\Big ) \theta\Big (|\mu-\nu|-\sqrt{M^2+q^2}\Big )\nonumber\\
&=&\left (\int_{0}^\infty-\int^{\nu_5}_0\right )\frac{dq}{\pi}\Big (|\mu-\nu|-\sqrt{M^2+q^2}\Big ) \theta\Big (|\mu-\nu|-\sqrt{M^2+q^2}\Big ),\label{D6}\\
V_2&=&\int_{-\nu_5}^\infty\frac{dq}{\pi}\Big (\mu+\nu-\sqrt{M^2+q^2}\Big ) \theta\Big (\mu+\nu-\sqrt{M^2+q^2}\Big )\nonumber\\
&=&\left (\int_{0}^\infty+\int^{\nu_5}_0\right )\frac{dq}{\pi}\Big (\mu+\nu-\sqrt{M^2+q^2}\Big ) \theta\Big (\mu+\nu-\sqrt{M^2+q^2}\Big ).\label{D7}
\end{eqnarray}
Due to the presence of the $\theta (x)$-function in the integrands of (\ref{D6}), the first integral there looks like
\begin{eqnarray}
\int_{0}^\infty \left (\cdots\right )\frac{dq}{\pi}&=&
\theta (|\mu-\nu|-M)\int_0^{\sqrt{|\mu-\nu|^2-M^2}}\frac{dq}{\pi}
\Big (|\mu-\nu|-\sqrt{M^2+q^2}\Big ),\label{D8}
\end{eqnarray}
whereas the second integral in this expression has the form
\begin{eqnarray}
\int_{0}^{\nu_5} \left (\cdots\right )\frac{dq}{\pi}&=&
\theta (|\mu-\nu|-M)\theta\left (\nu_5-\sqrt{|\mu-\nu|^2-M^2}\right )\int_0^{\sqrt{|\mu-\nu|^2-M^2}}\frac{dq}{\pi}
\Big (|\mu-\nu|-\sqrt{M^2+q^2}\Big )\nonumber\\
&+&\theta (|\mu-\nu|-M)\theta\left (\sqrt{|\mu-\nu|^2-M^2}-\nu_5\right )\int_0^{\nu_5}\frac{dq}{\pi}
\Big (|\mu-\nu|-\sqrt{M^2+q^2}\Big ).\label{D9}
\end{eqnarray}
Substituting the expressions (\ref{D8}) and (\ref{D9}) into (\ref{D6}) and using there the relation $\theta (x)=1-\theta (-x)$, we have
\begin{eqnarray}
V_1&=&
\theta (|\mu-\nu|-M)\theta\left (\sqrt{|\mu-\nu|^2-M^2}-\nu_5\right )\int^{\sqrt{|\mu-\nu|^2-M^2}}_{\nu_5}\frac{dq}{\pi}
\Big (|\mu-\nu|-\sqrt{M^2+q^2}\Big ).\label{D10}
\end{eqnarray}
In a similar way one can transform the expression (\ref{D7}) for $V_2$,
\begin{eqnarray}
V_2&=& 2\theta (\mu+\nu-M)\int^{\sqrt{(\mu+\nu)^2-M^2}}_{0}\frac{dq}{\pi}
\Big (\mu+\nu-\sqrt{M^2+q^2}\Big )\nonumber\\
&-&\theta (\mu+\nu-M)\theta\left (\sqrt{(\mu+\nu)^2-M^2}-\nu_5\right )\int^{\sqrt{(\mu+\nu)^2-M^2}}_{\nu_5}\frac{dq}{\pi}
\Big (\mu+\nu-\sqrt{M^2+q^2}\Big ).\label{D11}
\end{eqnarray}
Performing direct integrations in (\ref{D10}) and (\ref{D11}),
(recalling that $V=V_1+V_2$) and taking into account the relations (\ref{D1}) and
(\ref{D2}), then completes the derivation of formula (\ref{33}). By analogy, one can derive the expression (\ref{34}).

\end{document}